\documentclass[%
12pt,
reprint,
onecolumn,
tightenlines,
superscriptaddress,
notitlepage,
preprintnumbers,
nofootinbib,
 amsmath,amssymb,amsthm,
 aps,
eqsecnum,
]{revtex4-1}

\usepackage{isomath}
\usepackage{amsmath}
\usepackage{amsbsy}
\usepackage{amssymb}
\usepackage{amscd}
\usepackage{amsfonts}
\usepackage{stmaryrd}
\usepackage{siunitx}
\usepackage{euscript}
\usepackage[utf8]{inputenc}
\usepackage[T1]{fontenc}
\usepackage{newtxtext} 
\everymath{\displaystyle}
\usepackage{exscale}

\usepackage{graphicx}
\usepackage{boxedminipage}
\usepackage{calc}
\usepackage[usenames,dvipsnames]{xcolor}
\graphicspath{ {media/} }
\usepackage[caption=false]{subfig}

\usepackage[small,compact]{titlesec}
\usepackage{setspace}
\usepackage{enumitem}
\setitemize{noitemsep,topsep=0pt,parsep=0pt,partopsep=0pt}
\setenumerate{noitemsep,topsep=0pt,parsep=0pt,partopsep=0pt}
\setdescription{noitemsep,topsep=0pt,parsep=0pt,partopsep=0pt}

\usepackage[colorinlistoftodos, color=green!40,prependcaption]{todonotes}

\usepackage[,colorlinks,urlcolor=blue,citecolor=blue]{hyperref}
\usepackage[normalem]{ulem}

\makeatletter
\renewcommand{\p@subsection}{}
\renewcommand{\p@subsubsection}{}
\makeatother

\usepackage{isomath}
\usepackage{amsmath}
\usepackage{amssymb}
\usepackage{amscd}
\usepackage{amsfonts}

\newcommand{\Z}{\mathbb{Z}}

\newcommand{\half}{\frac{1}{2}}

\DeclareMathOperator{\GCD}{GCD}

\newcommand{\dm}{\ \mathrm{d}}

\newcommand{\bfa}{{\mathbold a}}

\newcommand{\bfc}{{\mathbold c}}

\newcommand{\bfe}{{\mathbold e}}

\newcommand{\bfk}{{\mathbold k}}

\newcommand{\bfn}{{\mathbold n}}

\newcommand{\bft}{{\mathbold t}}

\newcommand{\bfx}{{\mathbold x}}

\newcommand{\bfF}{{\mathbold F}}

\newcommand{\bfI}{{\mathbold I}}

\newcommand{\bfQ}{{\mathbold Q}}
\newcommand{\bfR}{{\mathbold R}}

\usepackage[plain]{fancyref}

\begin{document}


\preprint{To appear in Mathematics and Mechanics of Solids (\href{https://doi.org/10.1177/1081286520961831}{doi:10.1177/1081286520961831})}

\title{Symmetry-Adapted Tight-Binding Electronic Structure Analysis of Carbon Nanotubes with \\ Defects, Kinks, Twist, and Stretch}

\author{Soumya Mukherjee}
    \affiliation{Department of Civil and Environmental Engineering, Carnegie Mellon University}

\author{Hossein Pourmatin}
    \affiliation{Department of Civil and Environmental Engineering, Carnegie Mellon University}

\author{Yang Wang}%
    \affiliation{Pittsburgh Supercomputing Center}%
    \affiliation{Pittsburgh Quantum Institute, University of Pittsburgh}

\author{Timothy Breitzman}
    \affiliation{Air Force Research Laboratory}
    
\author{Kaushik Dayal}
    \email{Kaushik.Dayal@cmu.edu}
    \affiliation{Pittsburgh Quantum Institute, University of Pittsburgh}
    \affiliation{Center for Nonlinear Analysis, Department of Mathematical Sciences, Carnegie Mellon University}
	\affiliation{Department of Materials Science and Engineering, Carnegie Mellon University}
    \affiliation{Department of Civil and Environmental Engineering, Carnegie Mellon University}
    
\date{\today}


\begin{abstract}
    This paper applies a symmetry-adapted method to examine the influence of deformation and defects on the electronic structure and band structure in carbon nanotubes.
	First, the symmetry-adapted approach is used to develop the analog of Bloch waves.  
	Building on this, the technique of perfectly-matched layers is applied to develop a method to truncate the computational domain of electronic structure calculations without spurious size effects. 
	This provides an efficient and accurate numerical approach to compute the electronic structure and electromechanics of defects in nanotubes.
	The computational method is applied to study the effect of twist, stretch, and bending, with and without various types of defects, on the band structure of nanotubes.  
	Specifically, the effect of stretch and twist on band structure in defect-free conducting and semiconducting nanotubes is examined, and the interaction with vacancy defects is elucidated. 
	Next, the effect of localized bending or kinking on the electronic structure is studied.  
	Finally, the paper examines the effect of 5-8-5 Stone-Wales defects.  
	In all of these settings, the perfectly-matched layer method enables the calculation of localized non-propagating defect modes with energies in the bandgap of the defect-free nanotube.
\end{abstract}

\maketitle


\section{Introduction}

The electronic properties of nanotubes has been a focus of research for over two decades due to the richness of the physics and potential applications, e.g. \cite{PhysRevB.67.161401, bekyarova2005electronic, zhao2004electronic,bandaru2005novel,charlier2007electronic,javey2009carbon,bradley2003flexible,mceuen2002single,dekker2018we,cao2019review,wang2013carbon}.
An important feature of nanotubes is that their electronic properties have been found to be very sensitive to the chirality of the nanotube, e.g. \cite{kramberger2003assignment,avouris2003carbon,Chen2007228,papadopoulos2000electronic,PhysRevLett.94.156802}. 
The aim of this paper is to develop a symmetry-adapted electronic structure method, based on tight-binding, to examine the behavior of nanotubes that are uniformly deformed.
While this has been studied by other groups, notably including \cite{zhang2009electromechanical,zhang2008elasticity,nikiforov2014tight,kumar2020bending,ahmadpoor2017thermal,zelisko2017determining}, we build further on the symmetry-adapted approach to develop a method to study defects such as vacancies, Stone-Wales defects, and geometric defects such as localized bending or kinking.
Our approach to study defects in a computationally tractable manner uses perfectly-matched layers (PML) \cite{Pourmatin2016115}, which have been used to truncate the computational domain for wave equations posed on large or unbounded domains in other contexts such as elastodynamics \cite{basu2004perfectly} and electromagnetism \cite{BERENGER1994185}.
Specifically, PML provides a strategy to truncate the domain by using dissipative layers near the boundary of the truncated domain.
By using an appropriate construction of the dissipative layers, there are no spurious reflections and standard Dirichlet boundary conditions can be applied on the outer boundary of the truncated domain.
Other methods to account for defects in electronic structure calculations -- in the setting of crystals -- include flexible boundary conditions \cite{tan2019dislocation} and real-space methods \cite{ghosh2017sparc}.
Field-theoretic methods have been useful in providing insights in the setting of graphene and nanotubes \cite{zhang2006atomistic,li2005continuum,ariza2010discrete,ariza2012stacking}.

Our symmetry-adapted approach builds on the Objective Structures (OS) framework introduced by James \cite{james2006objective}, and developed further by Dumitrica, James and others \cite{dumitricua2007objective,james2006objective,dayal2010nonequilibrium,aghaei2013symmetry,aghaei2013anomalous}.
The OS approach provides a simple and physical way to apply uniform deformations such as twist and stretch.
Further, using the OS framework, it can be demonstrated that the tight-binding Schrodinger equation satisfies an analog of the Bloch theorem in crystals, even when there is uniform twist and/or stretch.
Prior work has shown that, even in the defect-free case, stretching is expected to lead to significant changes in the electronic properties \cite{yang2000electronic}.
Theoretical and experimental work have also shown that defects, such as Stone-Wales defects, can have unusual effects on the electronic structure \cite{robertson2012spatial,wang2015large}.
In this work, we study a combination of these settings.
Specifically, we study the influence of stretch and twist on conducting and semiconducting chiral and armchair nanotubes in the defect-free and defected settings; we examine the effect of localized bending or kinking on the defect modes; and we examine the electronic structure of $5-8-5$ Stone-Wales defects.

\textbf{Organization.}
In \Fref{sec:OS}, we describe the symmetry-adapted geometric description for nanotubes with twist, stretch or bending.
In \Fref{sec:TBM}, we describe the tight-binding method and parameters; the analog of Bloch waves in nanotubes; and the consequent method of perfectly matched layers to study defects.
In \Fref{sec:twisted_nano}, we apply the method to study the electronic structure of nanotubes under torsion and stretch in defect-free nanotubes and those with a vacancy defect.
In \Fref{sec:bent_nano}, we apply the method to a nanotube with localized bending or a kink.
Finally, in \Fref{sec:SW}, we examine a $5-8-5$ Stone-Wales triplet.


\section{Symmetry-Adapted Description of Carbon Nanotubes}
\label{sec:OS}

A compact and practical symmetry-adapted description of carbon nanotubes is provided by the framework of Objective Structures (OS) \cite{james2006objective}.
We build on this framework to develop a geometric description of deformed nanotubes.

Using the theory of isometry groups in 3D following \cite{dayalFormulae,dayal2010nonequilibrium,aghaei2013symmetry,aghaei2011symmetry,aghaei2013symmetry}, a discrete group $G$ of isometries consists of elements of the form $g = (\bfQ|\bfc)$ where $\bfQ$ is an orthogonal transformation and $\bfc$ is a vector.
The action of these isometries on any point $\bfx$ is given by $g(\bfx) = \bfQ \bfx + \bfc$.

We use the action of an isometry on a point to infer the successive action of group elements using composition of mappings:
\begin{equation}
g_1(g_2(\bfx)) = \bfQ_1 (\bfQ_2 \bfx + \bfc_2) + \bfc_1 = \bfQ_1 \bfQ_2 \bfx  + \bfQ_1 \bfc_2 + \bfc_1 = g_1 g_2(\bfx).
\end{equation}
This provides a natural definition for the group operation: given any 2 group elements $g_1 = (\bfQ_1|\bfc_1)$ and $g_2 = (\bfQ_2|\bfc_2)$, the group operation is $g_1 g_2 = (\bfQ_1 \bfQ_2| \bfQ_1 \bfc_2 + \bfc_1)$.
The group operation also enables us to define the identity $Id = (\bfI| 0)$ and inverses $g^{-1} = (\bfQ^T|-\bfQ^T\bfc)$.

We now apply the group-theoretic framework to a nanostructure that can be described by the OS framework.
Consider a unit cell with $M$ atoms, and label the positions $\bfx_{0,k}, k=1,\ldots,M$ where the subscript $0$ denotes the simulated unit cell.
The non-simulated images are obtained using the formula:
\begin{equation}
\bfx_{i,k}=g_i(\bfx_{0,k})=\bfQ_i \bfx_{0,k} +\bfc_i, \quad g_i\in G, \quad k=1,2,\ldots M \label{eqn:OS_example}
\end{equation}
Given an isometry group $G=\{g_0:=Id,g_1,\ldots,g_{N-1}\}, g_i=(\bfQ_i | \bfc_i)$, with $N$ being the number of image unit cells that could be either infinite or finite, the formula above gives us a description of the entire structure.

Specializing this to single-wall carbon nanotubes, following \cite{dayal2010nonequilibrium}, we consider a nanotube with chirality described by $(n,m)$, and nanotube axis $\bfe$.
Define the integers $p$ and $q$ such that $pm_1-qn_1=1$, where $m_1 =\frac{m}{\GCD(n,m)}, n_1 =\frac{n}{\GCD(n,m)}$ and $\GCD(n,m)$ is the greatest common divisor of $n$ and $m$.
We can then define the {\em generators} $h_1, h_2$ of this nanotube as isometries of the form:
\begin{equation}
\label{eqn:nanotube-parameters}
\begin{split}
	h_1 & = \left(\bfR_{\theta_1}|\left(\bfI-\bfR_{\theta_1}\right)\bfa\right), \quad \bfR_{\theta_1} \bfe = \bfe, \quad 0<\theta_1=\frac{2\pi \min \left(|p|,|q|\right)}{\GCD(n,m)} \leq 2\pi 
	\\
	h_2 &=\left(\bfR_{\theta_2}|\left(\bfI-\bfR_{\theta_2}\right)\bfa+\kappa_2 \bfe\right), \ \bfR_{\theta_2}\bfe = \bfe, \ \theta_2=\pi \frac{p(2n+m)+q(n+2m)}{n^2+m^2+nm}, \ \kappa_2=\frac{3\GCD(n,m)}{2 \sqrt{n^2+m^2+ n m}}l_0
\end{split}
\end{equation}
The generator $h_1$ is a rotation operation with axis coinciding with $\bfe$, and $h_2$ is a screw operation with the orthogonal part having the axis coinciding with $\bfe$. 
The quantity $l_0=0.142$ $\mathrm{nm}$ is the bond length of the graphite sheet before rolling.
The radius of the nanotube is given by $r_0=\frac{l_0}{2\pi}\sqrt{3(m^2+n^2+mn)}$.

The vector $\bfa$ defines the spatial location of the axis of the nanotube.
That is, the axis of the nanotube, oriented along $\bfe$, passes through the point $\bfa$.
To see this, we notice the action of $h_1$ and $h_2$ on a point $\bfx$:
\begin{equation}
\begin{split}    
h_1(\bfx)
& =
\bfR_{\theta_1}\bfx+\left(\bfI-\bfR_{\theta_1}\right)\bfa
\\
h_2(\bfx)
& =
\bfR_{\theta_2}\bfx+\left(\bfI-\bfR_{\theta_2}\right)\bfa + \kappa_2 \bfe
\end{split}    
\end{equation}
If $\bfx$ lies on the axis of the nanotube, i.e. $\bfx$ has the form $\bfa+\alpha\bfe$ where $\alpha$ is any real number, we find that it is mapped to a point on the axis:
\begin{equation}
\begin{split}    
h_1(\bfa+\alpha\bfe)
& =
\bfa+\alpha\bfe
\\
h_2(\bfa+\alpha\bfe)
& =
\bfa+\alpha\bfe + \kappa_2 \bfe
\end{split}    
\end{equation}
Hence, the line $\bfa+\alpha\bfe$ is mapped to itself, defining the axis.

The isometry group $G$ has elements $h_1^{i_1} h_2^{i_2}$ where $i_1$ and $i_2$ go over all integers. As shown in \cite{dayal2010nonequilibrium}, $G$ provides an isometry group to generate a nanotube that can be written:
\begin{equation}
\bfx_{(i_1,i_2),k}=h_1^{i_1} h_2^{i_2} (\bfx_{(0,0),k}), \quad i_1,i_2 \in\Z; k=1,2
\end{equation}
In this description, the unit cell contains two carbon atoms and hence $k=1,2$.
The parameters $\bfx_{(0,0),k}$ are listed in \cite{dumitricua2007objective}.

We note certain features of the OS isometry group approach.
First, OS above enables facile MD calculations of nanotubes that are chiral whereas periodic boundary conditions would require extremely long unit cells.
Imposed twist can also break translational symmetry completely, and would require, in principle, infinitely long unit cells with periodic MD, but can be easily and efficiently handled with OS.
Further, OS enables a transparent approach to the application of external loads that cause twisting and extension: the choices for $h_1, h_2$ in \eqref{eqn:nanotube-parameters} correspond to an unloaded nanotube; changing $\theta_2$ would correspond to imposed twist, while changing $\kappa_2$ would correspond to imposed extension, as we describe in the next section.

\subsection{Nanotubes with Stretching, Twisting and Bending}
\label{sec:nstb}

The OS framework provides a simple strategy to impose twisting (torsion) and stretch, e.g. \cite{aghaei2013symmetry,aghaei2011symmetry,zhang2008elasticity}.

Twist is imposed by changing the value of $\theta_2$ for the screw transformation $h_2$ to a value different from the equilibrium value shown in \eqref{eqn:nanotube-parameters}. 
That is, to impose a twist angle of $\tau$ per unit length, the value of $\theta_2$ is increased by $\tau \kappa_2$, where $\kappa_2$ is the axial translation  associated with $h_2$ described in \eqref{eqn:nanotube-parameters}.

Stretch is similarly imposed by changing the value of $\kappa_2$ away from the equilibrium value listed in \eqref{eqn:nanotube-parameters}.
That is, to impose a stretch $\lambda$, we replace $\kappa_2$ by $\lambda\kappa_2$ in \eqref{eqn:nanotube-parameters}.

Combined stretch and twist is imposed simply by a combination of both of these operations.

We next consider bending.
Uniform bending can be readily imposed by using a different symmetry-adapted description and changing the parameters as in the case of twist and stretch \cite{dumitricua2007objective}, and this would distort the nanotube away from equilibrium but retain a uniform structure.
In this paper, we instead consider localized bending motivated by experimentally observed configurations.
The nanotube is largely straight and uniform, except for a localized region in which bending deformation occurs in a kink-like manner.
We construct such geometries by changing the group parameters in \eqref{eqn:nanotube-parameters} to go between a uniform nanotube with axis $\bfe_1$ and a uniform nanotube with axis $\bfe_2$, with a gradual variation between these uniform end-states.

Consider a bend localized between the atoms numbered $N_1$ and $N_2$ along the helix. 
The axis of the nanotube is assumed to change linearly between $\bfe_1$ and $\bfe_2$ between these atoms:
\begin{equation}
\bfe(i) =  \frac{\frac{N_2 - i}{N_2-N_1} \bfe_1 +  \frac{i - N_1}{N_2-N_1} \bfe_2}{\left| \frac{N_2 - i}{N_2-N_1} \bfe_1 +  \frac{i - N_1}{N_2-N_1} \bfe_2 \right|}, \quad \forall \ i \in (N_1,N_2)
\label{eqn:bending-axis}
\end{equation}
Similarly, we use $\bfa(i) = i\kappa_2\bfe_1$ for $i \leq N_1$, and assume that it varies in the bend as $\bfa(i)= \bfa(i-1)+ \kappa_2\bfe(i)$.

The variation of $\bfe$ implies a corresponding variation in $\bfR_{\theta_1} (i)$ and $\bfR_{\theta_2}(i)$.
We keep $\theta_1$ and $\theta_2$ unchanged, and change the axis of the rotation $\bfe(i)$ using \eqref{eqn:bending-axis}.

The generators from \eqref{eqn:nanotube-parameters} will vary now vary along the nanotube as follows:
\begin{equation}
\label{eqn:nanotube-parameters_bending}
\begin{split}
h_1(i) & = \left(\bfR_{\theta_1}(i)| \left(\bfI-\bfR_{\theta_1}(i)\right)\bfa(i)\right), \quad \bfR_{\theta_1}(i) \bfe(i) = \bfe(i),
\\
h_2(i) &=\left(\bfR_{\theta_2}(i)| \left(\bfI-\bfR_{\theta_2}(i)\right)\bfa(i)+\kappa_2 \bfe(i)\right),\quad \bfR_{\theta_2}(i)\bfe(i) = \bfe(i),
\end{split}
\end{equation}
where $i$ is the exponent of the helical transformation $h_2$. 
We notice that this no longer forms a group and the resulting structure is not an Objective Structure, but closely approximates one outside of the localized bending region.

With these variable generators, the atomic positions $x_{(i_1,i_2),k}$ can be written:
\begin{align}
	\bfx_{(0,i_2),k}&=h_2(i_2-1)\left(\bfx_{(0,i_2-1),k}\right),\\
	\bfx_{(i_1,i_2),k}&=h_1^{i_1}(i_2-1)\left(\bfx_{(0,i_2-1),k}\right).\
\end{align}
The generators now provide an operation to position an atom given the position of the previous atom, but since it is not a group, the order of operation cannot be changed; however, this is not simply a non-Abelian group because it violates the closure property of a group.
Examples of bent nanotube geometries that can be generated with this formula are shown in Figure \ref{fig:bent-nanotube-1}.


\section{Tight Binding Analysis of Nanotubes}
\label{sec:TBM}

In this section, we describe the key details of the tight binding approach (\Fref{sec:tb-formulation}), the application to uniformly twisted and stretched nanotubes using the analogies between OS and perfect crystals (\Fref{sec:tb-perfect}), and then the application to bent nanotubes using perfectly matched layers (\Fref{sec:PML_theory}).

\subsection{Tight-Binding Formulation}
\label{sec:tb-formulation}

We start from the 1-electron Schrodinger wave equation:
\begin{equation}
    -\half\nabla^2\phi(\bfx) + V(\bfx)\phi(\bfx)=E\phi(\bfx),\label{Schrodinger11}\\
\end{equation}
where $\phi(\bfx)$ is the electron wave function and $V(\bfx)$ is the electrostatic potential. 
The tight-binding model decomposes the wave function in terms of the atomic orbitals \cite{tadmor2011modeling}:
\begin{equation}
    \phi(\bfx) = \sum_{i_1,i_2,k,\alpha}{c_{i_1 i_2 k\alpha}\phi_\alpha\left(\bfx-\bfx_{(i_1,i_2),k}\right),} 
    \label{LCAO1}
\end{equation}
where $\phi_{\alpha}$ is the orbital shape function for atomic orbital $\alpha$, and the indices $i_1,i_2$ and $k$ index the atoms in the manner described in \Fref{sec:OS}.

All carbon atoms are $sp^2$ hybridized in nanotubes, as one $2s$ orbital together with the $2p_x$ and $2p_y$ orbitals generate three $sp^2$ orbitals. 
Each $sp^2$ orbitals forms $\sigma$ bonds with the $sp^2$ orbitals of the neighboring carbon atoms. 
The one remaining electron per carbon atom in the $p_z$ orbital moves out of the plane to form $\pi$ bonds with the neighbouring $2p_z$ orbitals.
The energy states associated with the in-plane $\sigma$ bonds do not influence the electronic properties of nanotubes \cite{charlier2007electronic}.
However, the $\pi$ orbital energy states cross the Fermi energy level and hence play a key role in nanotube electronics \cite{charlier2007electronic,wallace1947band}. 
The change in configuration of nanotube due to torsion, bending etc., changes the relative orientation of these $p_z$ orbitals forming $\pi$ bonds which in turn alters its electronic properties \cite{zhang2009electromechanical,zhang2008elasticity}.
Hence, only the $p_z$ orbitals are used for the tight-binding model in this paper.
We adopt Gaussian type shape functions provided in equation (2.3) in \cite{hehre1969self} to numerically represent these orbitals.

\subsection{Twisting and Stretching of Nanotubes without Defects: Bloch Wave Analogs}
\label{sec:tb-perfect}

Objective Structures have numerous analogies to crystals, as described broadly in, e.g., \cite{james2006objective,aghaei2013symmetry,dumitricua2007objective,aghaei2011symmetry,dayal2010nonequilibrium,aghaei2012tension}.
These analogies have been utilized to study mechanical \cite{aghaei2012tension} and electromechanical \cite{Popov_2004,zhang2009electromechanical} behavior of nanotubes under stretch and torsion and to examine stability \cite{zhang2008stability} and defects \cite{zhang2009dislocation} of nanostructure under torsion.
The key application of this analogy here is that we are able to restrict our calculations to a small repeating unit cell.
In crystals, this is possible due to the classical Bloch theorem, e.g. \cite{martin2004electronic}.
Here, we describe the analog of the Bloch theorem in the nanotube setting using the symmetry-adapted Objective Structure formulation.

Consider a perfect nanotube, that is, infinitely long and straight with no defects.
The electrostatic potential $V_p(\bfx)$ in this setting has the same symmetry as the nanotube:
\begin{equation}
    V_p\left(g^i(\boldsymbol{x})\right)=V_p(\boldsymbol{x}), \forall i\in \mathbb{Z}
    \label{vp}
\end{equation}
where $g$ is any element of the isometry group, and $i$ runs over the set of integers $\mathbb{Z}$.

Using the 1-electron Schrodinger model, we have:
\begin{equation}
    -\half\nabla^2\phi(\bfx) + V_p(\bfx)\phi(\bfx)
    =
    E\phi(\bfx),\label{Schrodinger1}\\
\end{equation}
We show below, following the classical Bloch theorem, that the wave function $\phi(x)$ does {\em not} inherit the symmetry of $V_p$ from \eqref{vp}, but instead picks up a phase factor.

Following the general structure of the classical Bloch theorem \cite{martin2004electronic}, we begin by introducing two functionals $\Pi_1$ and $\Pi_2$, which operate on any function $\chi(\bfx)$ as $\Pi_1\left[\chi(\bfx)\right]=\chi\left(h_1(\bfx)\right)$ and $\Pi_2\left[\chi(\bfx)\right]=\chi\left(h_2(\bfx)\right)$.
We notice that $\Pi_1^i\left[\chi(\bfx)\right]=\chi\left(h_1^i(\bfx)\right)$ and $\Pi_2^i\left[\chi(\bfx)\right]=\chi\left(h_2^i(\bfx)\right)$, where the superscript denotes the exponent defined as composition of mappings.

Considering the action of $\Pi_1$ on the Hamiltonian $H$, we have:
\begin{equation}
    \Pi_1 H\phi(\boldsymbol{x})
    = 
    \Pi_1\left[\left(-\frac{1}{2}\frac{\partial^2}{\partial x_i\partial x_i}+V_p(\boldsymbol{x})\right)\phi(\boldsymbol{x})\right]
    =
    \left(-\frac{1}{2}\frac{\partial^2}{\partial \left(h_1(\bfx)\right)_i\partial \left(h_1(\bfx)\right)_i} +V_p(h_1(\boldsymbol{x}))\right)\phi(h_1(\boldsymbol{x}))
    \label{111_1}
\end{equation}
where subscripts denote components and repeated indices imply summation.
From \eqref{vp}, we have $V_p\left(h_1(\boldsymbol{x})\right)=V_p(\boldsymbol{x})$.
From \cite{james2006objective, dumitricua2007objective}, we have that:
\begin{equation}
    \frac{\partial\quad }{\partial \left(h_1(\bfx)\right)_i}=\left(\bfR_{\theta_1}\right)_{ij}\frac{\partial\,  }{\partial x_j}
    \Rightarrow
	\frac{\partial^2}{\partial \left(h_1(\bfx)\right)_i\partial \left(h_1(\bfx)\right)_i}
	=
	\left(\bfR_{\theta_1}\right)_{ij}\left(\bfR_{\theta_1}\right)_{il}\frac{\partial^2}{\partial x_j\partial x_l}
	=
	\frac{\partial^2}{\partial x_i\partial x_i}
    \label{eq:partial}
\end{equation}
where $\bfR_{\theta_1}$ is the orthogonal transformation corresponding to $h_1$, introduced in \eqref{eqn:nanotube-parameters}.
Therefore, we have:
\begin{equation}
    \Pi_1 H\phi(\boldsymbol{x})
    =
    \left[\left(-\frac{1}{2}\frac{\partial^2}{\partial x_i\partial x_i}+V_p(\boldsymbol{x})\right)\Pi_1\left[\phi(\boldsymbol{x})\right]\right]
    =
    H\Pi_1 \left[\phi(\boldsymbol{x})\right].
\end{equation} 
That is, $H$ and $\Pi_1$ commute.
Through an exactly analogous calculations, we can show that $H$ and $\Pi_2$ commute.

A consequence of the commutation relation is that $H$, $\Pi_1$ and $\Pi_2$ have a common set of eigenfunctions \cite{martin2004electronic}.
Hence, $\phi(\boldsymbol{x})$, an eigenfunction of $H$, is also an eigenfunction of $\Pi_1$ and $\Pi_2$:
\begin{align}
	\Pi_1\left[\phi(\boldsymbol{x})\right] = t_1\phi\left(\boldsymbol{x}\right) \Rightarrow \Pi_1^{i_1}\left[\phi(\boldsymbol{x})\right] =t_1^{i_1}\phi\left(\boldsymbol{x}\right), \forall i_1\in \mathbb{Z}
	\\
	\Pi_2\left[\phi(\boldsymbol{x})\right] = t_2\phi\left(\boldsymbol{x}\right) \Rightarrow \Pi_2^{i_2}\left[\phi(\boldsymbol{x})\right] =t_2^{i_2}\phi\left(\boldsymbol{x}\right), \forall i_2\in \mathbb{Z}
	\label{pii}
\end{align}
where $t_1$ and $t_2$ are the eigenvalues of $\Pi_1$ and $\Pi_2$ corresponding to the eigenfunction $\phi$.
It follows immediately that
\begin{equation}
	\Pi_1^{i_1}\Pi_2^{i_2}\left[\phi(\boldsymbol{x})\right] = \Pi_2^{i_2}\Pi_1^{i_1}\left[\phi(\boldsymbol{x})\right]=t_2^{i_2}t_1^{i_1}\phi\left(\boldsymbol{x}\right)
	\label{eq:2iso}
\end{equation}

We first consider $t_1$.  
Notice that $h_1^{N_1} = (\bfI | \bf0)$, where $N_1 = 2\pi / \theta_1$, i.e., if we apply the rotation operation $N_1$ times, we recover the identity operation.
Therefore:
\begin{equation}
	\Pi_1^{N_1}\left[\phi(\bfx)\right] = t_1^{N_1}\phi\left(\bfx\right) = \phi\left(\bfx\right) 
	\Rightarrow t_1^{N_1} = 1 
	\Rightarrow t_1 = e^{\imath \frac{2\pi}{N_1} n}, n \in \{ 0, 1, \ldots N_1-1\}
\end{equation}
Using $N_1 = 2\pi / \theta_1$ and defining $k_1 = n / r_0$, we can write 
\begin{equation}
	t_1 = e^{\imath k_1 \theta_1 r_0}, \quad k_1 r_0 \in \{ 0, 1, \ldots N_1-1\}
	\label{eqn:t1}
\end{equation}
The arbitrary quantity $k_1$ has the natural interpretation of an angular wavenumber.

We next consider $t_2$.
Unlike $h_1$, no power of the screw operation $h_2$ gives the identity operation, and all integer powers must be considered.
We first define $\gamma_2$ as the length between two consecutive atoms along the helix:
\begin{equation}
	\gamma_2=\sqrt{r_0^2\theta^2_2+\kappa_2^2}.
	\label{ga2}
\end{equation}
Consider the eigenvalues $T_2(i_2 \gamma_2), T_2(j_2 \gamma_2), T_2((i_2+j_2) \gamma_2)$ corresponding to the operations $\Pi_2^{i_2}, \Pi_2^{j_2}, \Pi_2^{i_2 + j_2}$ respectively; notice that these operations correspond to moving along the helix by a distance $i_2 \gamma_2, j_2 \gamma_2, (i_2+j_2) \gamma_2$ respectively.
Since it is equivalent whether we move in two succesive steps of $i_2 \gamma_2, j_2 \gamma_2$ or directly by $(i_2+j_2) \gamma_2$, we have that $T_2(i_2 \gamma_2) \times T_2(j_2 \gamma_2) = T_2((i_2+j_2) \gamma_2)$.
Therefore $T_2$ must be an exponential in its arguments; further, since $T_2(i_2 \gamma_2)$ must be bounded for all $i_2 \in \mathbb{Z}$, it must have unit magnitude.
Therefore, $T_2(i_2\gamma_2) = e^{\imath i_2 k_2 \gamma_2}$, giving:
\begin{equation}
	t_2 = e^{\imath k_2 \gamma_2}, \quad k_2 \gamma_2 \in [-\pi, \pi)
	\label{eqn:t2}
\end{equation}
The arbitrary quantity $k_2$ has the natural interpretation of a helical wavenumber.

By substituting \eqref{eqn:t1}, \eqref{eqn:t2} in \eqref{eq:2iso}, we obtain the Bloch theorem analog for nanotubes:
\begin{equation}
	\Pi_1^{i_1}\Pi_2^{i_2}\left[\phi(\bfx)\right]=\phi\left(h_1^{i_1}h_2^{i_2}(\bfx)\right)=e^{\imath i_1 k_1 r_0 \theta_1 + \imath i_2 k_2 \gamma_2}\phi(\bfx)\label{OSSSS}
\end{equation}
Using LCAO, we can rewrite the form \eqref{OSSSS} as
\begin{equation}
	{\phi}_{k_1,k_2}(\bfx)
	=
	\sum_{\alpha,m,j_1,j_2} c_{\alpha m} e^{\imath j_1 k_1 r_0 \theta_1+\imath j_2 k_2 \gamma_2}\phi_{\alpha}(\boldsymbol{x}-\bfx_{(j_1,j_2),m}),\label{eee}
\end{equation}
where $\alpha$ denotes the atomic orbital.
This can be substituted in \eqref{Schrodinger1} for numerical calculation of Bloch-wave analog solutions. 
The Bloch theorem analog developed here is applicable to infinite nanotubes subject to uniform stretch and/or torsion, and also provides the analog of band structures and density of states, which we apply to carbon nanotubes in \ref{ddos}.


\subsection{Nanotubes with Defects: Perfectly Matched Layers}
\label{sec:PML_theory}

The Bloch wave analogs developed in section \ref{sec:tb-perfect} are restricted to infinite nanotubes that are uniformly deformed but without defects.
To enable the study of defects, we build on the Bloch wave analogs and use the approach of perfectly matched layers (PML) following \cite{Pourmatin2016115}, using the complex coordinate stretching approach \cite{chew19943d}.
PML enables us to focus the computational effort in the neighborhood of the defect and truncate at some distance away, with the perfect matching preventing the formation of spurious reflections and size-effects due to the finite domain.

We recall the single electron Schrodinger wave equation given by
\begin{equation}
    -\half\nabla^2\phi(\bfx) + V(\bfx)\phi(\bfx)=E\phi(\bfx).\label{Schrodinger121}\\
\end{equation}
For perfect OS with no defects, the Bloch-wave analog can be written as:
\begin{equation}
    -\frac{1}{2} \nabla^2\phi_\bfk(\boldsymbol{x})+V_p(\boldsymbol{x})\phi_\bfk(\boldsymbol{x})=E_\bfk\phi_\bfk(\boldsymbol{x}).\label{Schrodinger}\\
\end{equation}
where $\bfk = (k_1, k_2)$ is the Bloch wavevector.

In the presence of a localized (i.e., non-extended) defect, the electrostatic potential $V(\bfx)$ is decomposed into $V_p(\bfx)+V_d(\bfx)$, where $V_p$ is the potential of the perfect structure and $V_d$ is the defect potential.
Similarly, we decompose the wave function into $\phi(\bfx) = \phi_\bfk(\bfx) + \psi(\bfx)$, where the first term corresponds to the incoming Bloch wave and is obtained from \eqref{Schrodinger}, and $\psi$ is the scattered wave.
Using this decomposition in \eqref{Schrodinger121}, we get:
\begin{align}
	-\frac{1}{2}\nabla^2\left(\phi_\bfk(\bfx)+\psi(\bfx)\right)
	+
	\left(V_d(\bfx)+V_p(\bfx)\right)\left(\phi_\bfk(\bfx)+\psi(\bfx)\right)
	=
	E\left(\phi_k(\bfx)+\psi(\bfx)\right).\label{2}
\end{align}
Noticing that the scattered wave $\psi(\bfx)$ must vanish at $|\bfx|\to\infty$, \eqref{2} reduces, at $|\bfx|\to\infty$, to:
\begin{equation}
    -\frac{1}{2}\nabla^2\phi_\bfk (\boldsymbol{x})+V_p(\boldsymbol{x})\phi_\bfk(\boldsymbol{x})
    =
    E\phi_\bfk(\boldsymbol{x}), \quad \text{at } |\bfx|\to\infty
    .\label{Schrodinger19}
\end{equation}
Comparing \eqref{Schrodinger19} with \eqref{Schrodinger}, we deduce that $E=E_\bfk$; this is to be expected as the scattering is elastic and, therefore, no energy is lost.
Using that $E=E_\bfk$, we can subtract \eqref{Schrodinger} from \eqref{2} to obtain:
\begin{equation}
    -\frac{1}{2}\nabla^2\psi(\boldsymbol{x})+\left[V_d(\boldsymbol{x})+V_p(\boldsymbol{x})\right]\psi(\boldsymbol{x})+V_d(\boldsymbol{x})\phi_\bfk(\boldsymbol{x})=E_\bfk \psi(\boldsymbol{x}).\label{3}
\end{equation}
We notice that this is {\em not} an eigenvalue problem.
Rather, we first solve \eqref{Schrodinger} as an eigenvalue problem to find the energy $E_\bfk$ and wavefunction $\phi_\bfk$ of the incident wave given $\bfk$ using the Bloch wave approach developed in \Fref{sec:tb-perfect}, and then use that information in \eqref{3} to solve a standard linear problem for $\psi$.

To solve \eqref{3} numerically, we develop the weak form
\begin{equation}
\begin{split}
	\int_{\Omega} \left[\half 
	    \nabla \psi^* \cdot \nabla \psi 
	    +
	    \psi^*(\boldsymbol{x})\left(V_p(\boldsymbol{x})+V_d(\boldsymbol{x})-E_\bfk\right)\psi(\boldsymbol{x})
	    \right] 
    \dm V
    -
    \int_{\Gamma} \half \psi^*(\boldsymbol{x})\nabla\psi(\boldsymbol{x}) \cdot\hat\bfn\dm s
    \\
    =-\int_{\Omega_d}\psi^*(\boldsymbol{x})V_d(\boldsymbol{x})\phi_\bfk(\boldsymbol{x})\dm V
    .\label{4}
\end{split} 
\end{equation}
In this equation, $\psi^*$ is the complex conjugate of $\psi$; $\Omega_d$ is the bounded region where $V_d$ is nonzero; $\Omega$ is the computational domain that includes the defect as well as the surrounding region; and $\Gamma$ is the boundary, with outward unit normal $\hat\bfn$, of $\Omega$.
If $\Omega$ is taken to be sufficiently large, we can apply Dirichlet BCs on $\Gamma$ as the scattered wave would decay, but this would require a very large computational expense.
Instead, we solve \eqref{3} in a relatively small domain near the defect, and construct a thin perfectly matched layer in the vicinity of the boundary.
PMLs cause the scattered wave to decay quickly without spurious reflections into the interior of the domain, and we can safely apply Dirichlet BCs on $\Gamma$.

We use a fictitious coordinate transformation such that each point $\bfx$ in real space is mapped to $\tilde\bfx$:
\begin{equation}
    \tilde{\bfx}=\bfx+i\sigma(\boldsymbol{x}).\label{doit}
\end{equation}
The real smooth function $\sigma(\boldsymbol{x})$ is chosen to be $0$ outside the PML, and chosen to have the form \cite{Pourmatin2016115}:
\begin{equation}
    \sigma(\boldsymbol{x})
    =
    \sigma_0\frac{\int_{\bfx_0}^{\bfx}\left|(\bft-\bfx_0)\right|^2\left|(\bfx_1-\bft)\right|^2\mathrm{d}\bft}{\int_{\bfx_0}^{\bfx_1}\left|(\bft-\bfx_0)\right|^2\left|(\bfx_1-\bft)\right|^2\mathrm{d}\bft}
\end{equation}
within the PML.
Here, $\bfx_0$ is a point at the interface of the computational domain and the absorbing layer, and $\bfx_1$ is a point on the exterior boundary.
To accommodate this transformation, the weak form in \eqref{4} is written as
\begin{equation}
	\int_{\Omega}  \left[\half \psi_{,i}^*F_{ij}^{-1} \psi_{,n}F_{nj}^{-1} + \psi^*(\boldsymbol{x})\left(V_p(\boldsymbol{x})+V_d(\boldsymbol{x})-E_\bfk\right)\psi(\boldsymbol{x})\right] J \dm V
	=
	-\int_{\Omega_d}\psi^*(\boldsymbol{x})V_d(\boldsymbol{x})\phi_\bfk(\boldsymbol{x})J\dm V,\label{PML_eqn}\
\end{equation}
where repeated indices denote summation, $F_{ij}=\partial{\tilde{x}_{i}}/\partial{{x}_j}$, and $J=\det\bfF$.
The integral over $\Gamma$ vanishes due to the applied Dirichlet BCs.

The PML approach described above is useful to compute the scattering due to incident electrons.
However, defect modes, i.e. eigenstates of the Schrodinger equation that are in the bandgap of the perfect structure and hence localized to the defect, are also of interest.
To compute defect modes, we directly solve \eqref{Schrodinger121} as an eigenvalue problem, within a truncated domain using PML, without decomposing the potential or wavefunction.


\section{Stretched and twisted nanotubes with defects}
\label{sec:twisted_nano}

We now apply the framework developed above to study the effect of twisting and stretching on nanotubes of different chiralities, both defect-free as well with a vacancy defect.
Bloch wave analogs are used to obtain band structures for homogeneously deformed nanotubes of various chiralities, while PML is utilized for analyzing the influence of vacancy. 
Introducing stretch, torsion, or changes in chirality induces significant changes in band-structure and electron density.


\subsection{Density of states for nanotubes of various chiralities under torsion and stretch}
\label{ddos}

The band structures reported here are obtained by solving the Bloch wave analogs. 
The density of states (DOS), $D(E)$, is calculated following \cite{pickard1999extrapolative}:
\begin{equation}
    D(E)=\frac{1}{\sqrt{2\pi}\sigma}\sum_i{e^{-\left(E-E_i\right)^2/2\sigma^2}},\label{eq:dos}
\end{equation}
corresponding to the energy state $E$, where the expression is summed over $i$ for each energy level $E_i$; $\sigma$ is a smoothing parameter, which is taken to be $2\times 10^{-3}$. 

\begin{figure}[htb!]
    \centering
    \includegraphics[width=.8\textwidth]{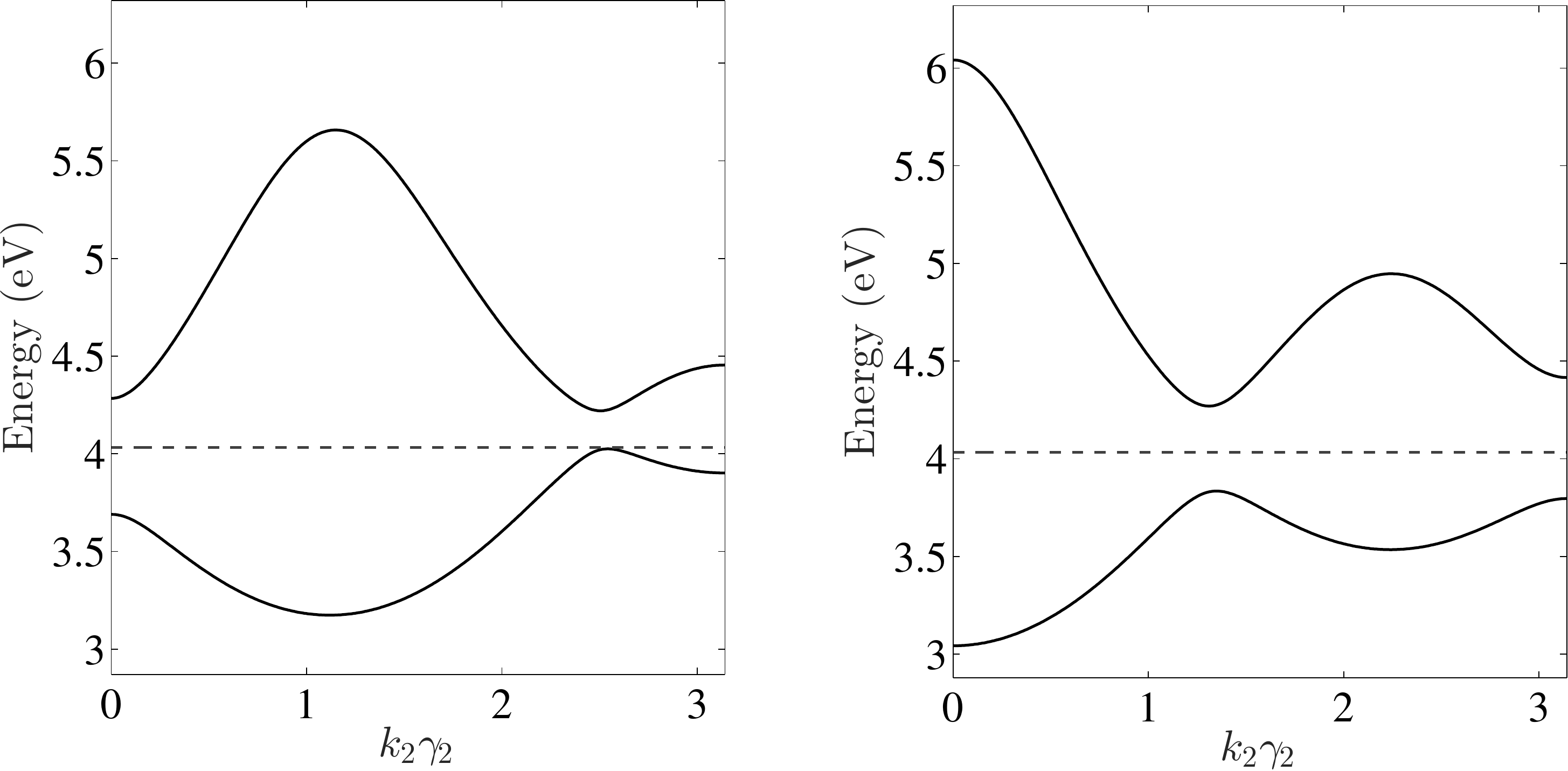} \
    \caption{Band structures of a (4,2) chiral nanotube, for $k_1=0$ (left) and $k_1=\frac{1}{r_0}$ (right). Fermi energy level is shown with a dotted line.}
    \label{bands}
\end{figure}

Both conducting and semiconducting nanotubes are probed under twist and stretch.
A chiral $(4,2)$ nanotube is used as a representative of chiral semiconducting nanotubes, while a $(5,5)$ armchair nanotube is used as a representative of conducting nanotubes.
Figure \ref{bands} shows the band structure for a $(4,2)$ chiral nanotube, selected for its semiconducting properties.
The energy is plotted against $k_2$, for several representative values of $k_1$.
Since $N_1=2$ for a $(4,2)$ carbon nanotube, the band-structure is plotted for $2$ independent values $k_1$ following \eqref{eqn:t1}.
As expected in $(4,2)$ nanotubes, we observe a bandgap.

\begin{figure}[htb!]
	\includegraphics[scale=.46]{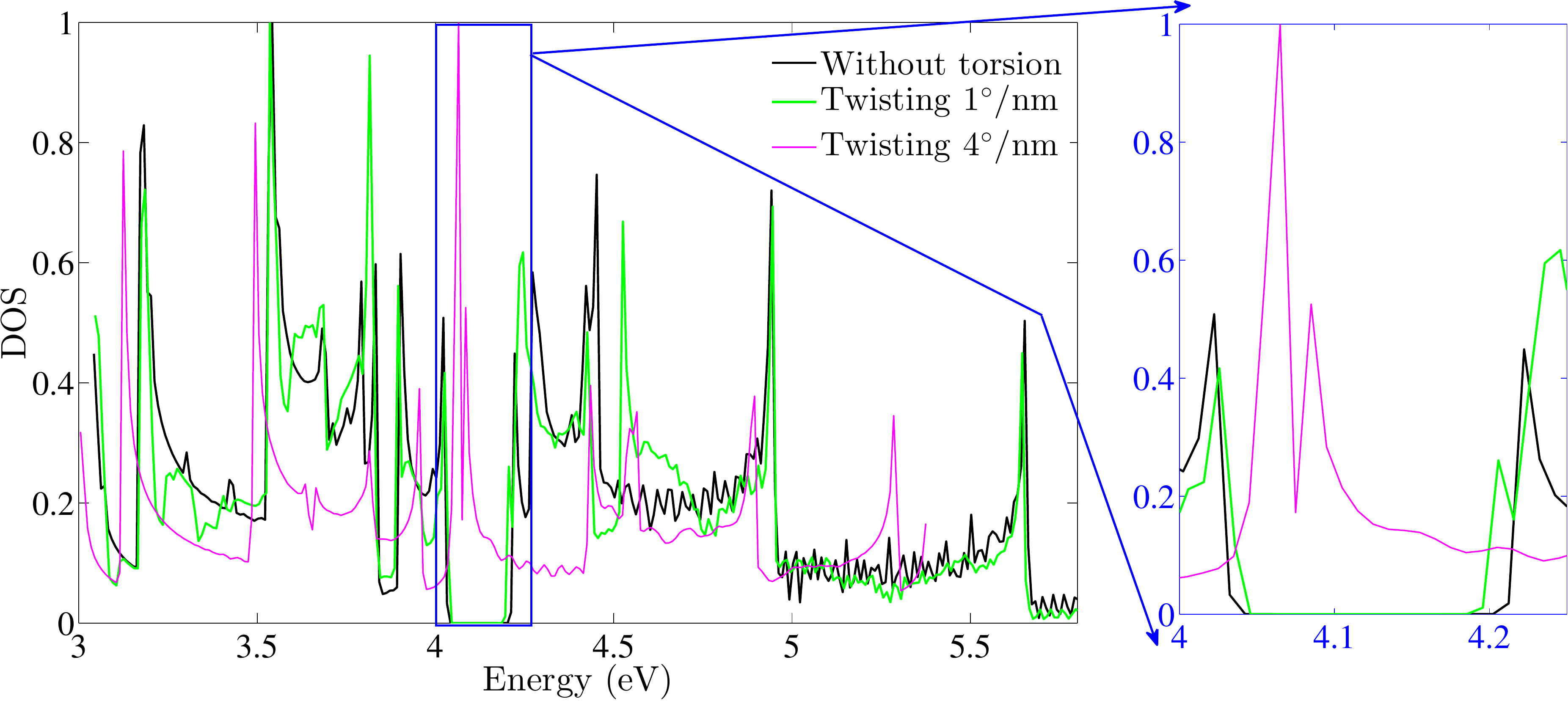} \\ 
	\includegraphics[scale=.5]{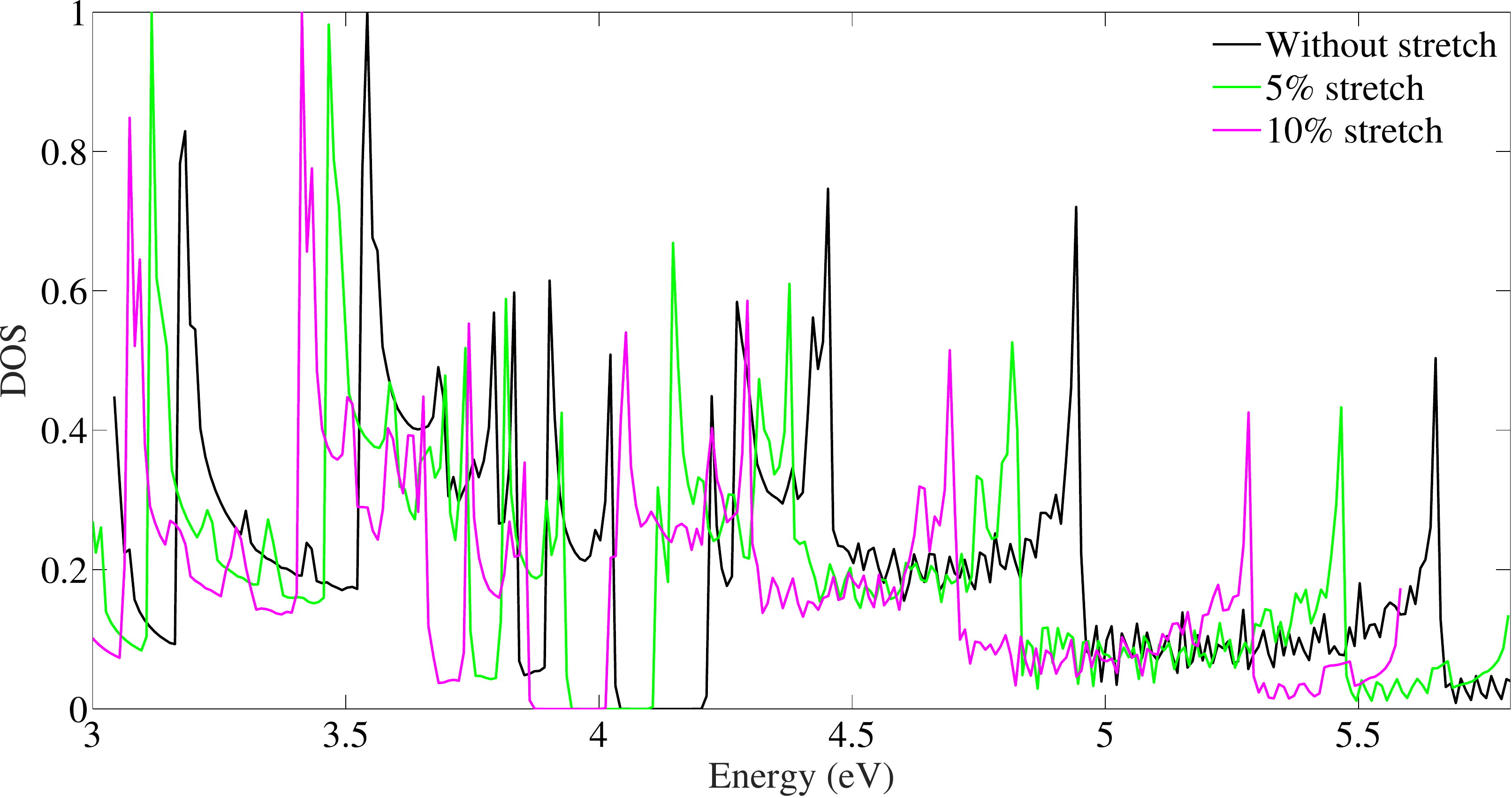}\ 
	\caption{DOS for $(4,2)$ CNT under varied amount of torsion (top) and stretch (bottom), with bandgap highlighted for torsion.}
	\label{DOS}
\end{figure}

\begin{figure}
	\includegraphics[scale=.57]{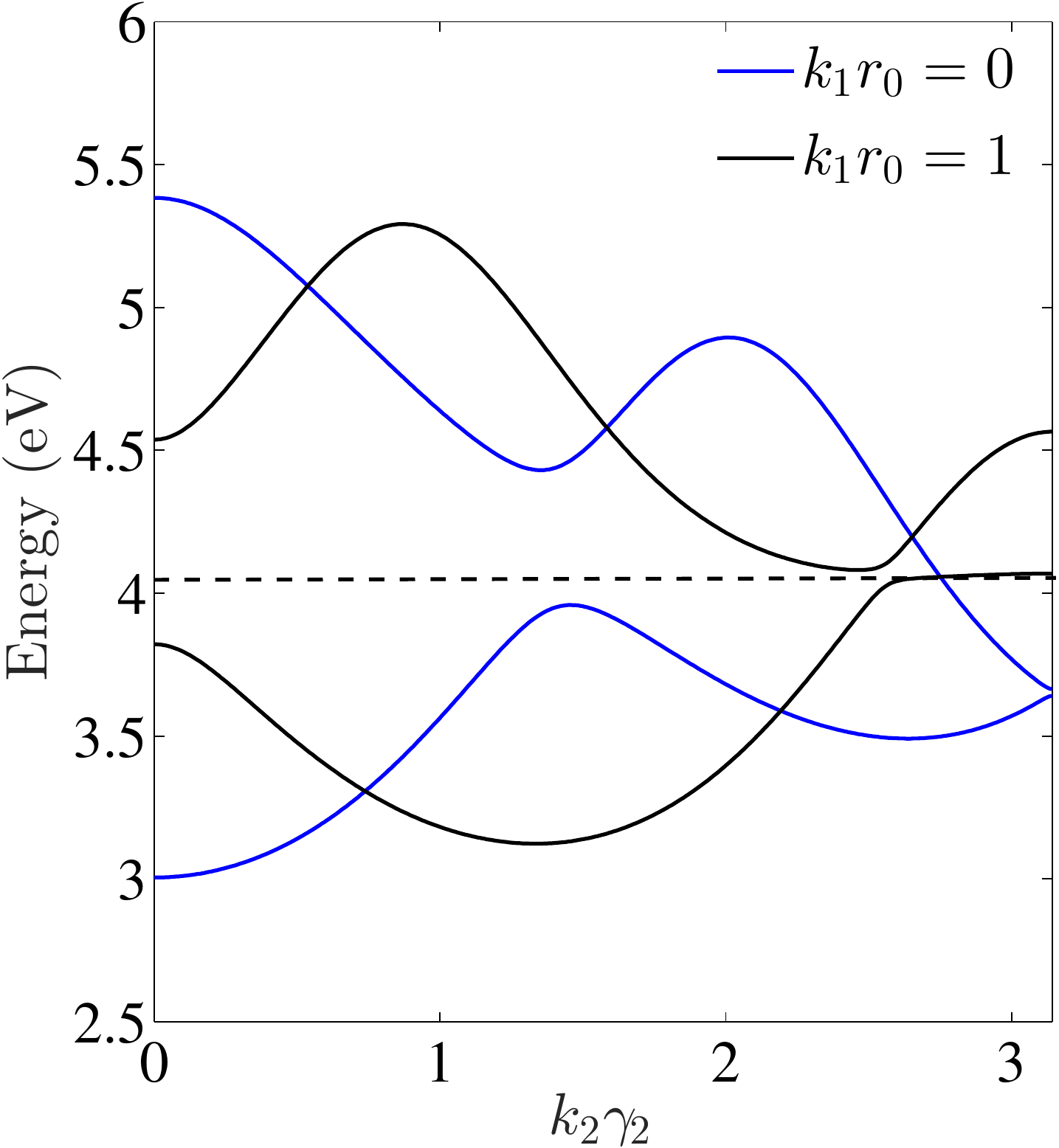}
	\caption{Band structure of (4,2) nanotube under $4^\circ/\mathrm{nm}$ twist for different values of $k_1$. Fermi energy level is shown with a dotted line.}
	\label{bands1}
\end{figure}

\begin{figure}[!htbp]
    \includegraphics[scale=.6]{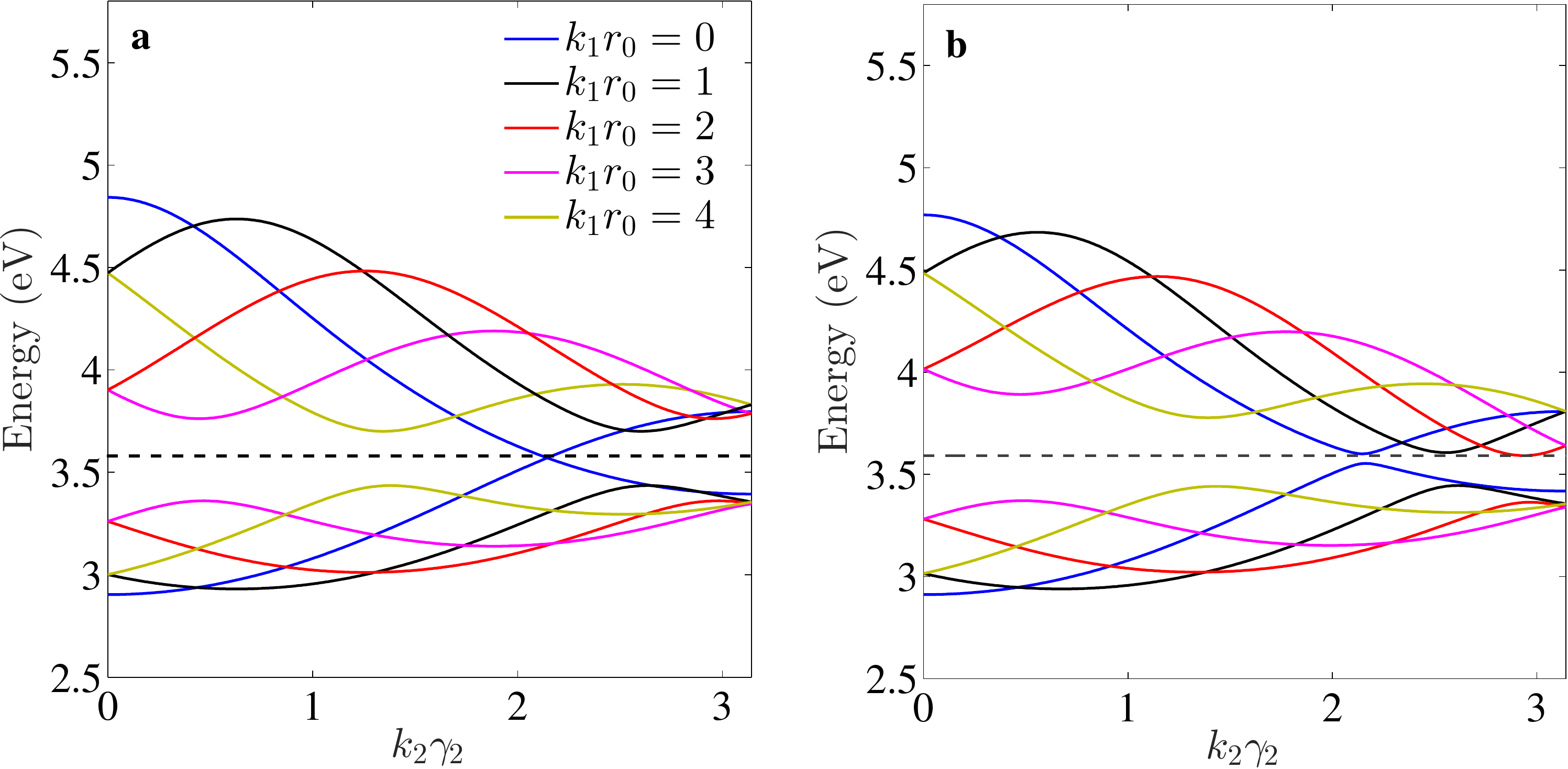}
	\caption{Band structures for a (5,5) armchair nanotube with (left) no twist and (right) twist of $\pm 2^\circ/\mathrm{nm}$. Fermi energy levels are shown with dotted lines.}
	\label{bands55}
\end{figure}

We then subject the nanotube to torsion and stretch.
Figure \ref{DOS} shows the density of states under different levels of torsion and stretch.
We observe a significant variation of electronic properties in Figure \ref{DOS} (top) under torsion for a $(4,2)$ chiral nanotube. 
We see that bandgap changes slightly for a twisting of $1^\circ/\mathrm{nm}$, but disappears completely at the twisting angle $4^\circ/\mathrm{nm}$; the corresponding band structure under a twist of $4^\circ/\mathrm{nm}$ is shown in in Figure \ref{bands1}.
However, significant qualitative changes are not observed due to stretching in Figure \ref{DOS} (bottom); roughly, the DOS simply translates in energy space.

Thus we observe that the bandgap of a semiconducting nanotube can disappear under sufficient levels of twist.
We next turn to the case of a $(5,5)$ conducting armchair nanotube, and find below that a bandgap can appear under sufficient levels of twist.

Figure \ref{bands55} shows the band structures of a $(5,5)$ nanotube with no twist and with twist of $2^\circ/\mathrm{nm}$ respectively with $N_1=5$ cases of $k_1$ for each case. 
We observe that a bandgap is introduced due to $\pm 2^\circ/\mathrm{nm}$ twist in the conducting $(5,5)$ armchair nanotube.
We point out that positive and negative twist provide atomic structures that are related by mirror symmetry, and hence have the same band structure.

This effect of twisting on band structure observed here agrees with the theoretical results given by \cite{yang2000electronic} and computational finding \cite{zhang2009electromechanical} on twisting of armchair nanotubes.
In general, we find that qualitative changes in the band structure can be induced by torsion.
These changes can be physically understood by noticing that twisting has a qualitatively similar effect to a change in chirality.
Relative rotation of $\pi$ orbitals induced in neighboring atoms due to twisting causes this change in electronic properties. 
On the other hand, we see that stretching causes fewer {\em qualitative} changes to the band structure.
Similar ideas have been investigated in the strain engineering of the graphene band structure using density functional theory \cite{kerszberg2015ab}, Green's function techniques \cite{sahalianov2019straintronics} and experimentation \cite{huang2011electronic}.


\subsection{Influence of vacancy defects}
\label{ppml}

We next study the influence of stretch, torsion, and chirality on the electron structure of nanotubes with vacancy defects.
When defects are present, the symmetry is broken, and hence we use the PML method described in section \ref{sec:PML_theory}.
We use this to obtain the electron density, and study the influence of torsion, stretch, and chirality on the electron density.

\begin{figure}[!http]
	\includegraphics[scale=.25]{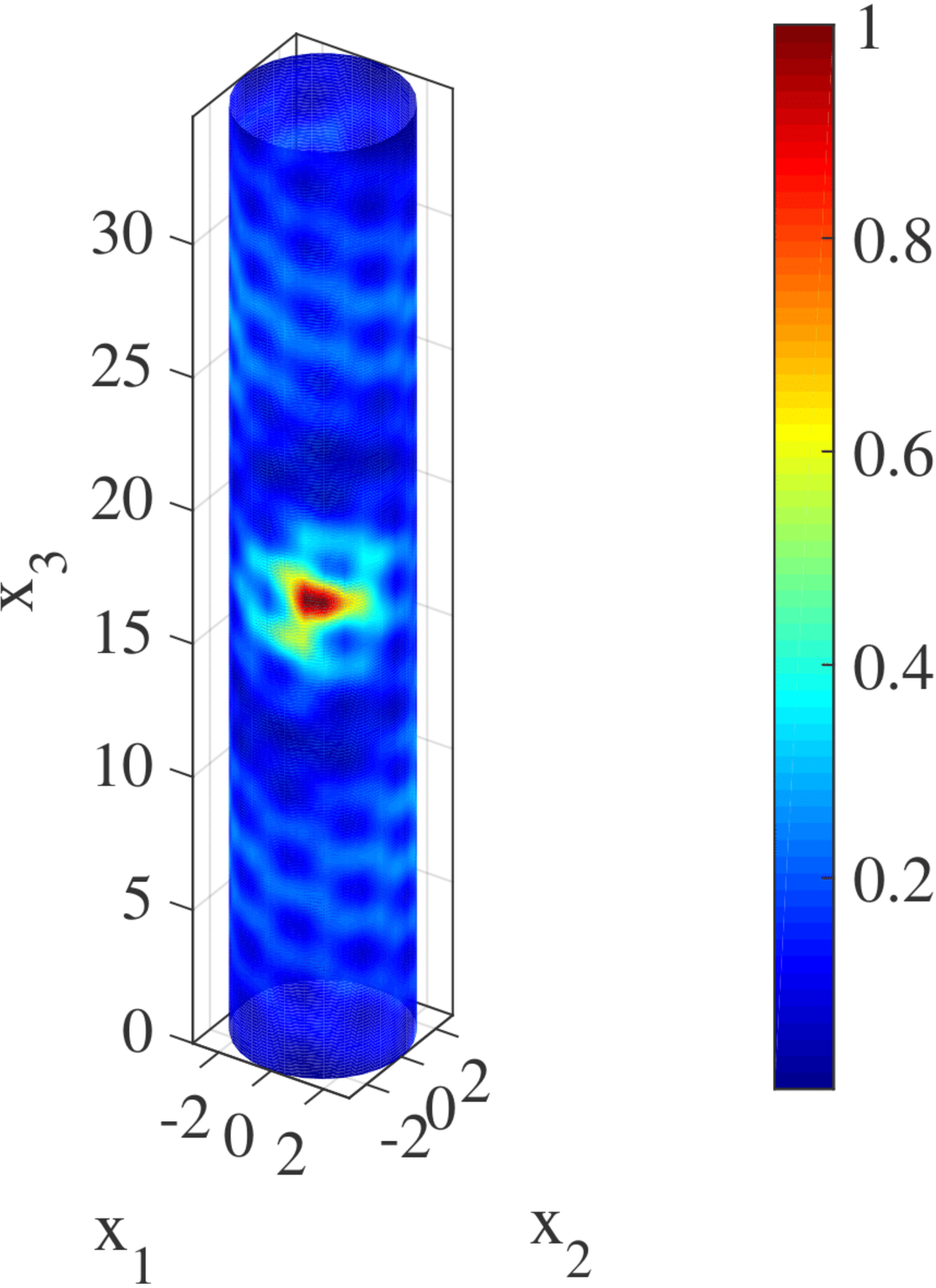} \
	\caption{Scattered electron density in a $(5,5)$ nanotube with vacancy defect with a twist of $3^\circ/\mathrm{nm}$. All lengths are in dimensions of $\mbox{\normalfont\AA}$.}
	\label{plot_torsion}
\end{figure}

\begin{figure}[!htbp]
	\includegraphics[width=0.47\textwidth]{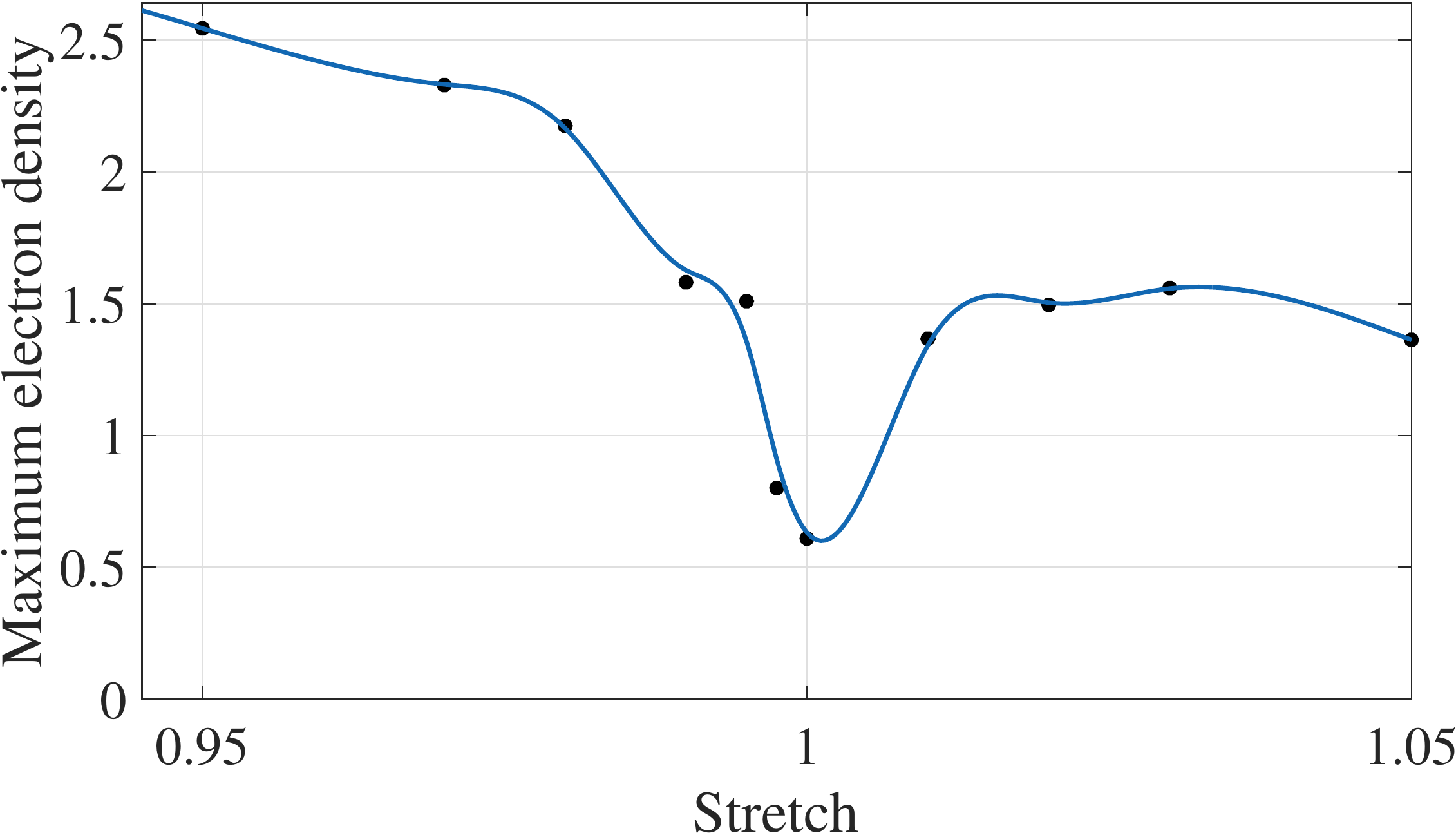}\
	\includegraphics[width=0.45\textwidth]{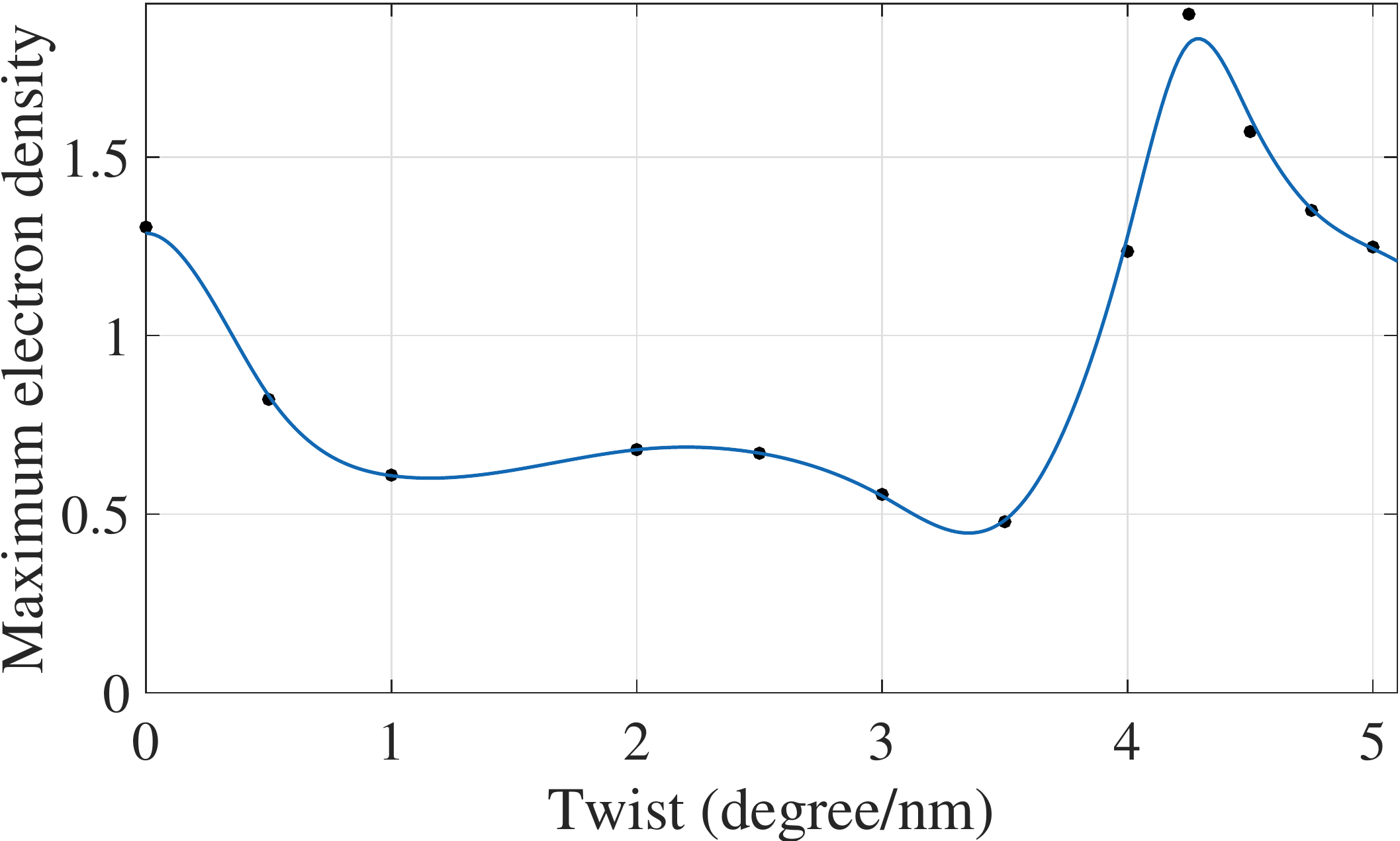}
	\caption{Influence of stretch and torsion on the maximum electron density. Left: Stretch is varied, with a constant pre-twist of $2^\circ/\mathrm{nm}$. Right: Twisting angle is varied. The lines are fits to the data points.}
	\label{torsion_plots}
\end{figure}

Figure \ref{plot_torsion} shows the scattered electron density for a twisted $(5,5)$ armchair nanotube with a vacancy defect subject to a twist of $3^\circ/\mathrm{nm}$.
To show the effect of stretch and torsion, Figure \ref{torsion_plots} plots the maximum electron density in the $(5,5)$ armchair nanotube with a vacancy for different levels of twist and stretch.
In Figure \ref{torsion_plots} (left), the level of stretch is varied with a constant pre-twist of $3^\circ/nm$, while Figure \ref{torsion_plots} (right) shows the effect of the twist angle. 
The continuous lines in Figure \ref{torsion_plots} are obtained by a curve fit.
Figure \ref{torsion_plots} (left) shows that both axial tension and compression have a tendency to increase the maximum electron density in comparison with the undeformed configuration. 
Torsion has a more complex effect on maximum electron density as observed in Figure \ref{torsion_plots} (right). 
We observe that while twisting, the maximum electron density gradually decreases initially with the increase of twist angle, followed by a sharp increase and then a decrease again. 
The twist angle was limited to $5^\circ/\mathrm{nm}$ to avoid torsional buckling at higher twist angles \cite{aghaei2012tension}.

\begin{figure}[htb!]
	\includegraphics[width=0.55\textwidth]{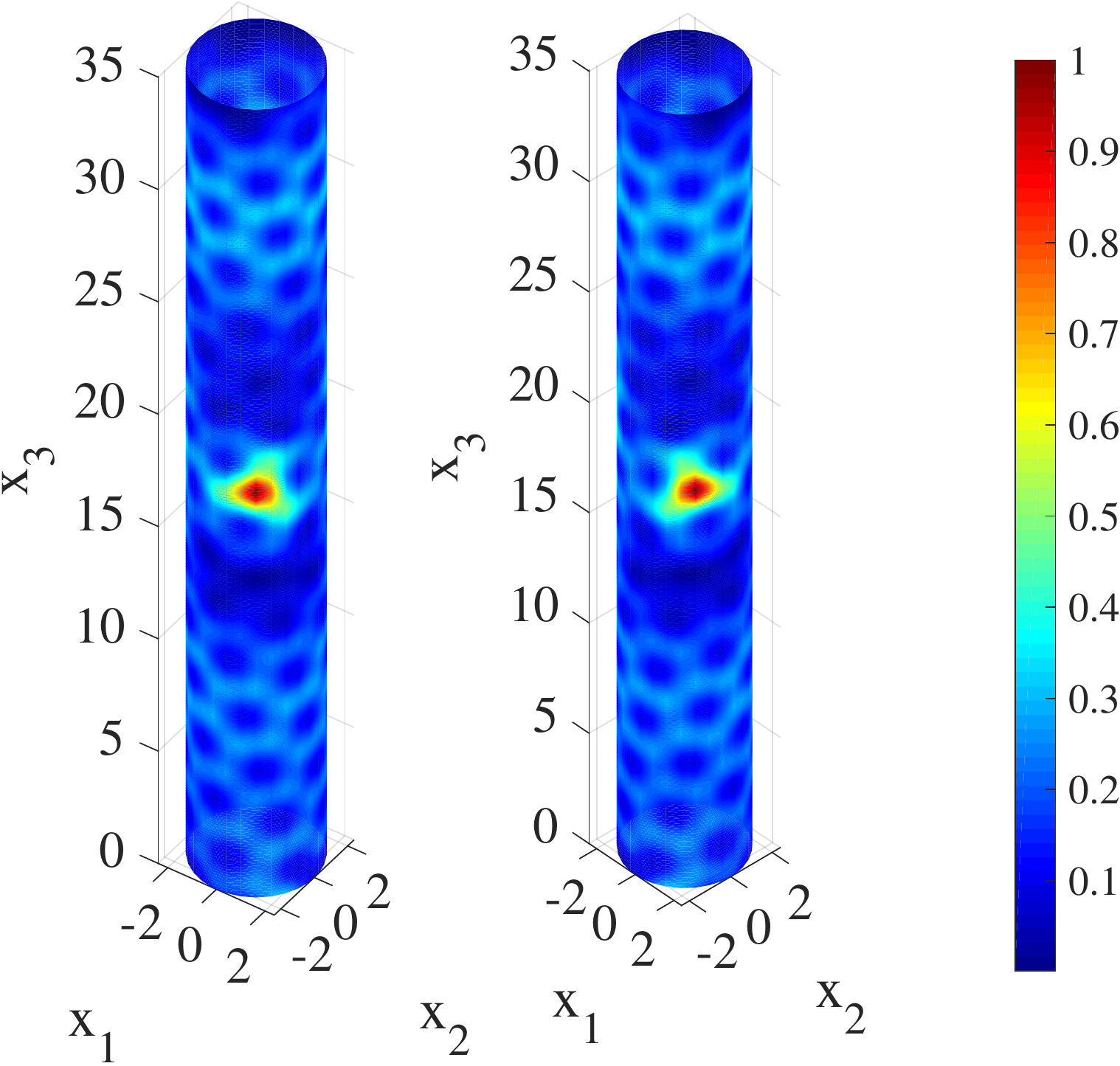}%
	\caption{Reflectional symmetry observed in the scattered electron density patterns of two mutually reflectionally symmetric nanotubes, for (2,5) on the left and (5,2) on the right, for the same incident Bloch wave of $k_1=0$ and ${k_2} = .347435$. All lengths are in dimensions of $\mbox{\normalfont\AA}$.}
	\label{chiral_plots}
\end{figure}

We next study the role of chirality of the nanotube.
Figure \ref{chiral_plots} show the electron density for $(2,5)$ and $(5,2)$ chiral nanotubes.
These nanotubes are chosen since they are examples of exact reflected copies of each other.
The electron density plots are mirror images of each other as these nanotubes are related by a reflection.

In Figures \ref{plot_torsion} and \ref{chiral_plots},  we also observe a repeating pattern in the electron density, except for the sharp jump near defect. 
This is caused by the dependence of the scattering wave function $\psi(\boldsymbol{x})$ on the Bloch-wave function $\phi_\bfk(\boldsymbol{x})$, as shown in \eqref{3}. 
The Bloch waves inherit the symmetry of the nanotube, which is passed on to the scattered wave function.


\section{Defect modes in kinked nanotubes}
\label{sec:bent_nano}

We next examine kinked nanotubes, with the geometric description from \Fref{sec:nstb}.
Unlike uniformly bent nanotubes, e.g. \cite{dumitricua2007objective}, kinked nanotubes must be considered as containing a geometric defect.
To obtain the defect modes, we use the PML method to focus on a region in the vicinity of the kink.
We focus on the chiral $(4,2)$ semiconducting nanotube and armchair $(5,5)$ conducting nanotube as in the previous section to enable comparisons.
Prior work \cite{shan2005first,meunier1998energetics} in seminconducting nanotubes has shown the possibility of additional defect modes in the bandgap due to bending. 

\begin{figure}[htb!]
	\label{fig:bent-nanotube-1}
	\includegraphics[scale=.35]{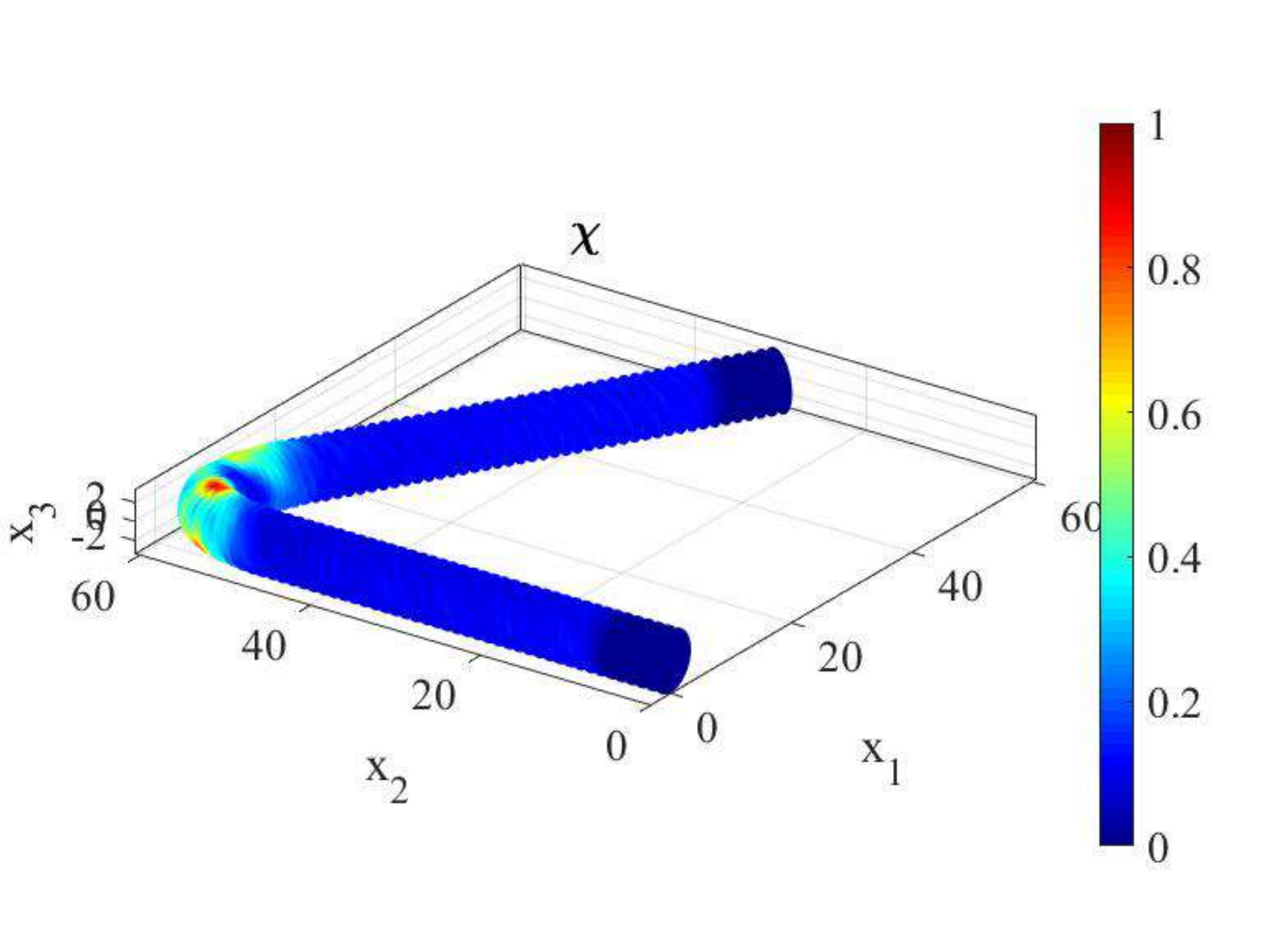}\
	\includegraphics[scale=.35]{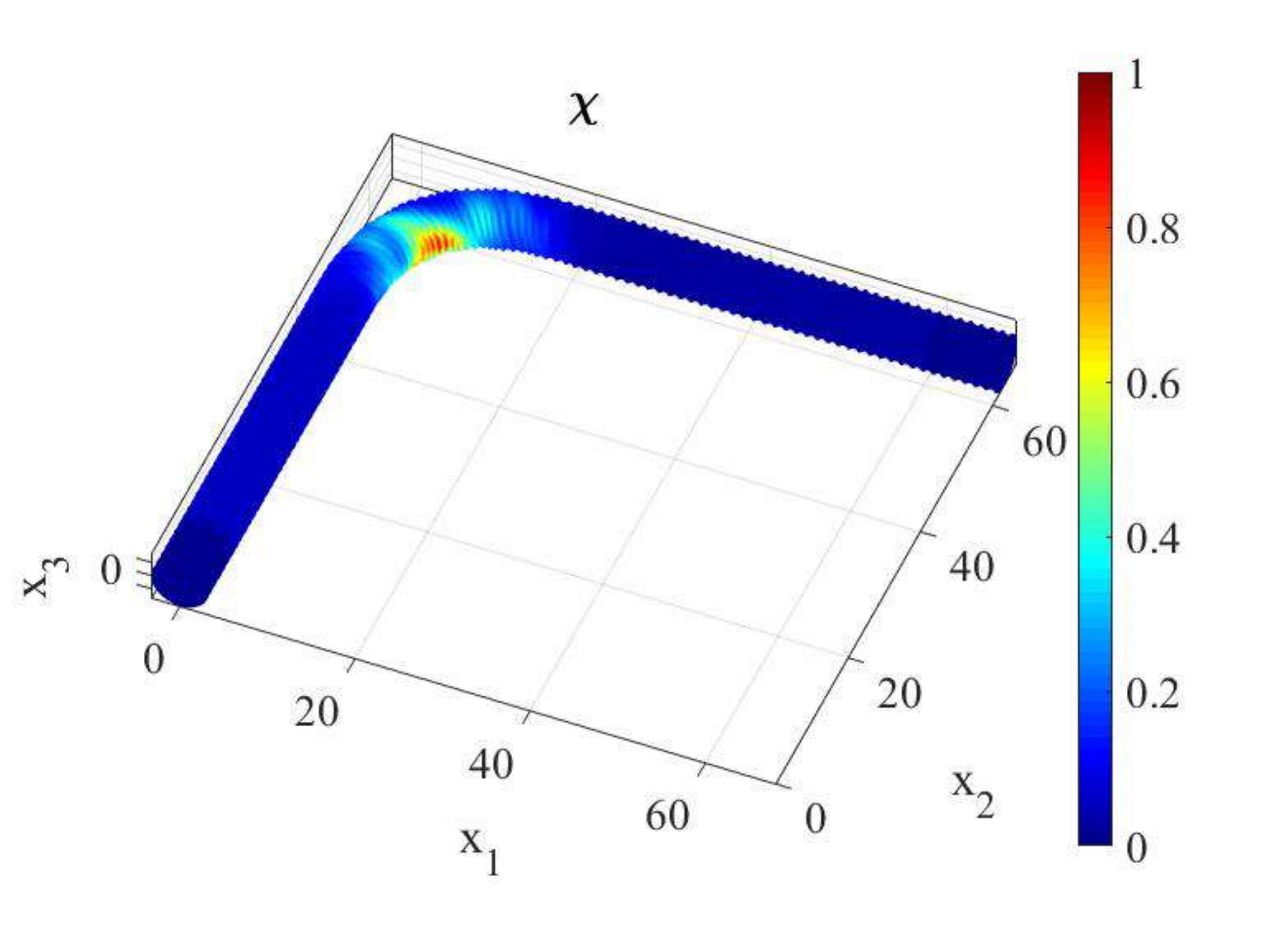}\
	\includegraphics[scale=.35]{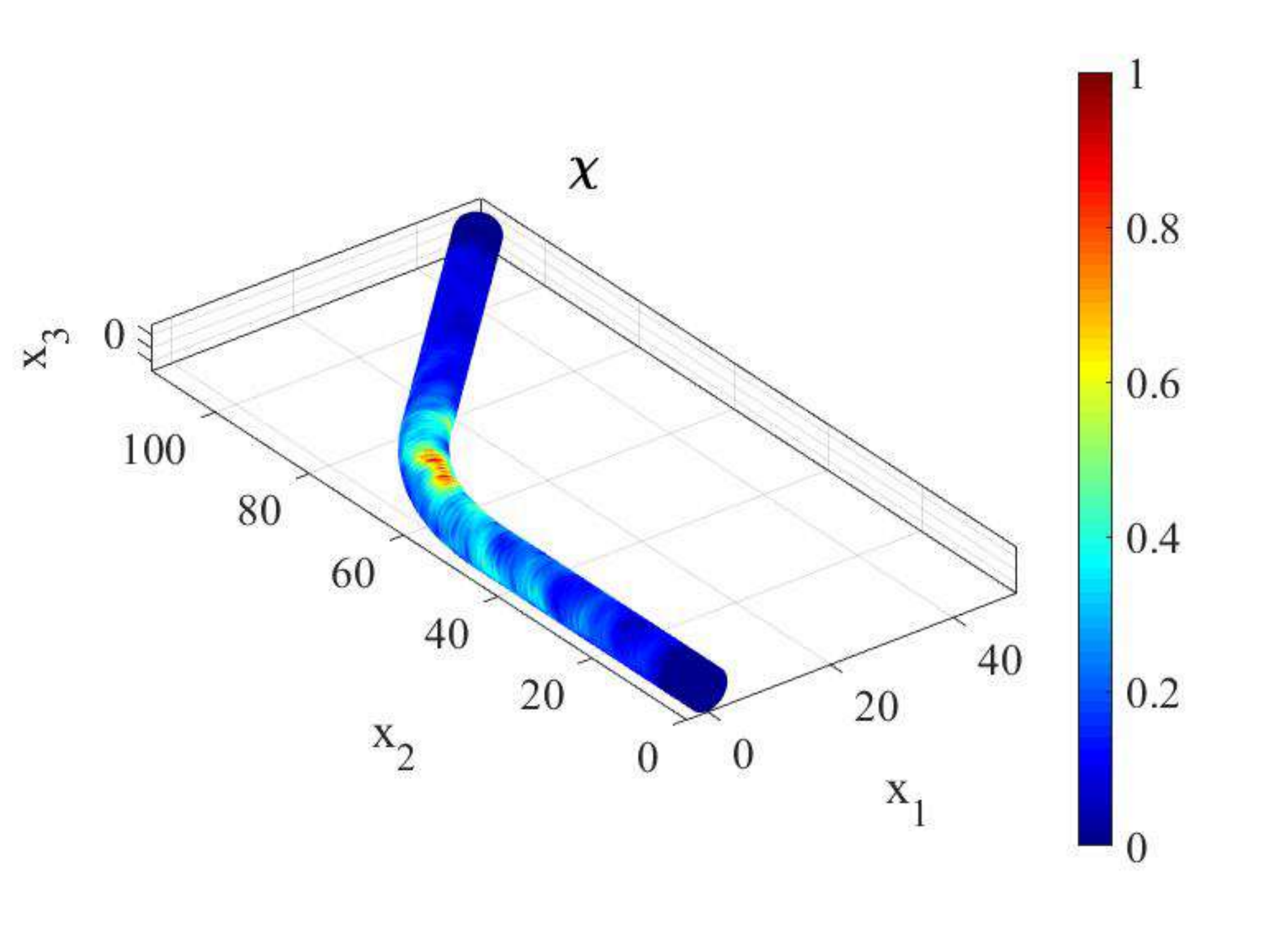}\
	\caption{A selected eigenstate, showing only the scattered wave, for bent $(5,5)$ nanotube, for different bending angles: (top left) $60^\circ$, (top right) $90^\circ$ and (bottom) $135^\circ$. All lengths are in dimensions of $\mbox{\normalfont\AA}$.}
	\label{bent-nanotube-1}
\end{figure}

In Figure \ref{bent-nanotube-1}, we examine the dependence on angle of kinking of a specific mode in a $(5,5)$ nanotube. 
We observe that the localization of the eigenstates increases with the sharpness of kinking.

\begin{figure}[htb!]
	\includegraphics[width=0.45\textwidth]{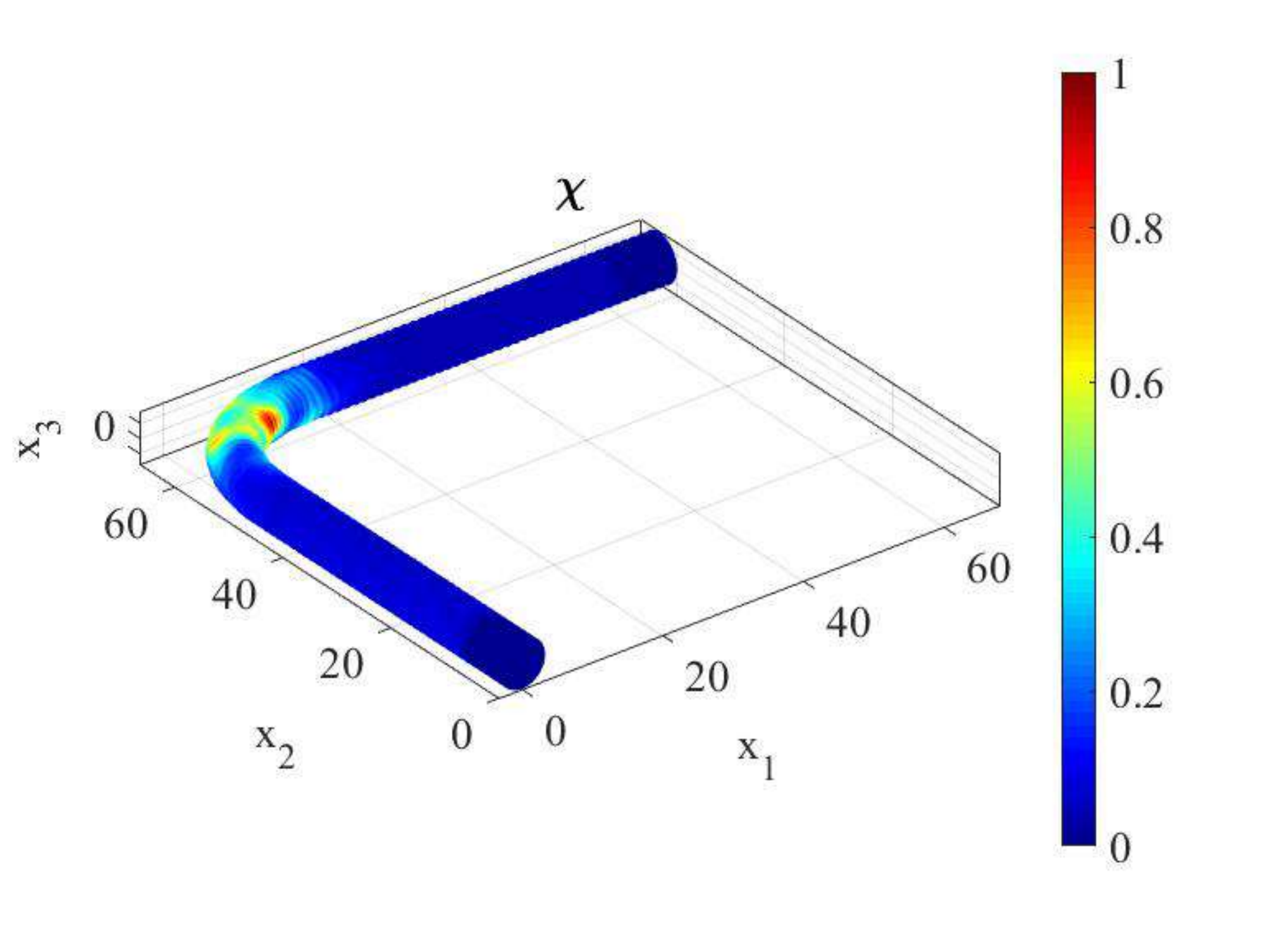}
	\
	\includegraphics[width=0.32\textwidth]{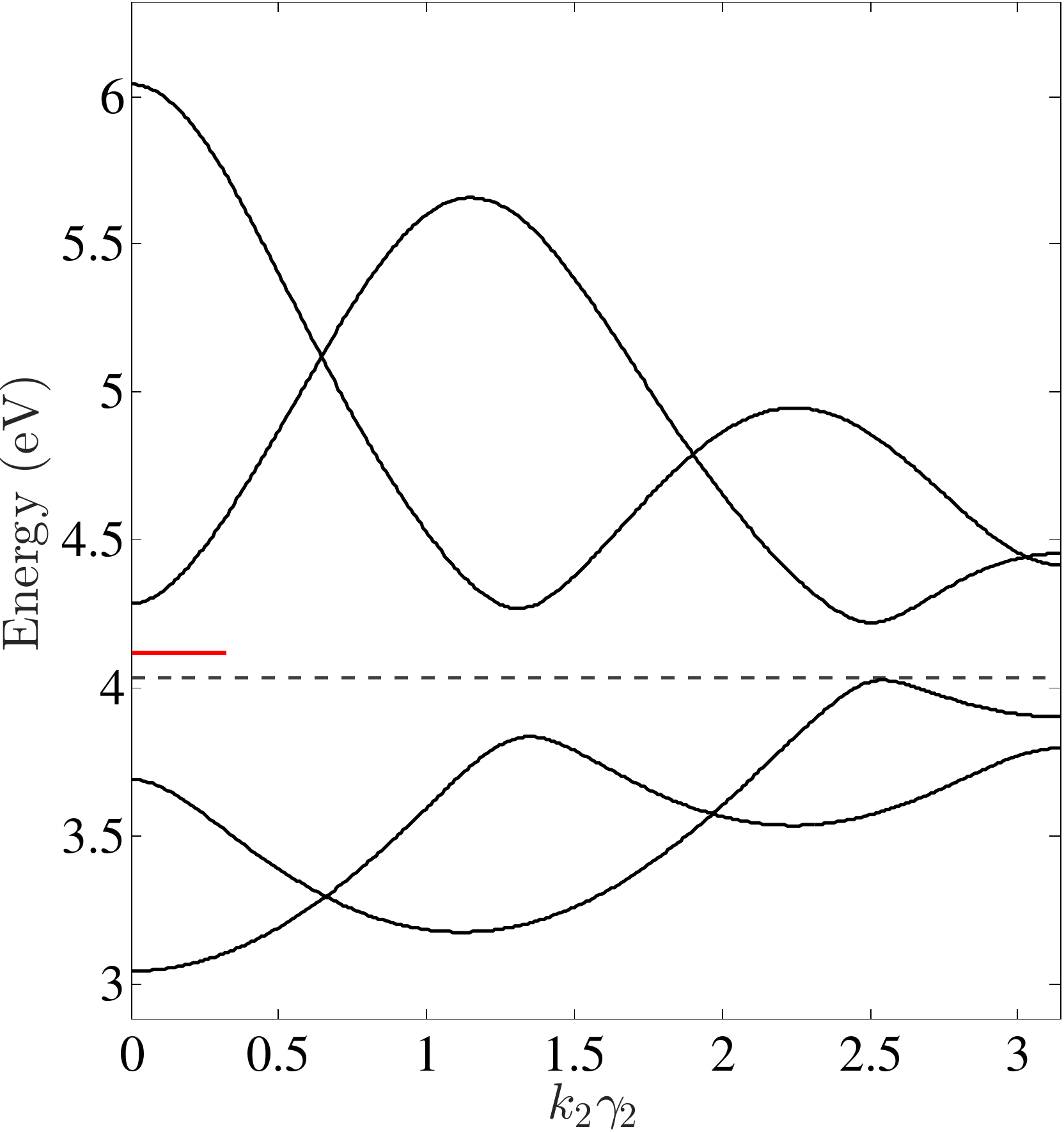}
	\caption{A selected defect eigenstate for $(4,2)$ carbon nanotube with the energy level in the bandgap: (left) electron density, and (right) band structure showing the defect mode energy (red line) in the bandgap. All lengths are in dimensions of $\mbox{\normalfont\AA}$.}
	\label{42}
\end{figure}

We next turn to the $(4,2)$ chiral nanotube that has a bandgap.
Figure \ref{42} shows a defect mode which corresponds to energy eigenvalue within the bandgap.
These additional modes within the bandgap are solely due to the defect \cite{shan2005first,meunier1998energetics,chico1996quantum,LAMBIN199585}. 
The PML-based approach enables us to compute the corresponding electronic structure. 


\section{Defect modes induced by a Stone-Wales defect}
\label{sec:SW}

\begin{figure}[htb!]
	\centering
	\includegraphics[width=0.55\textwidth]{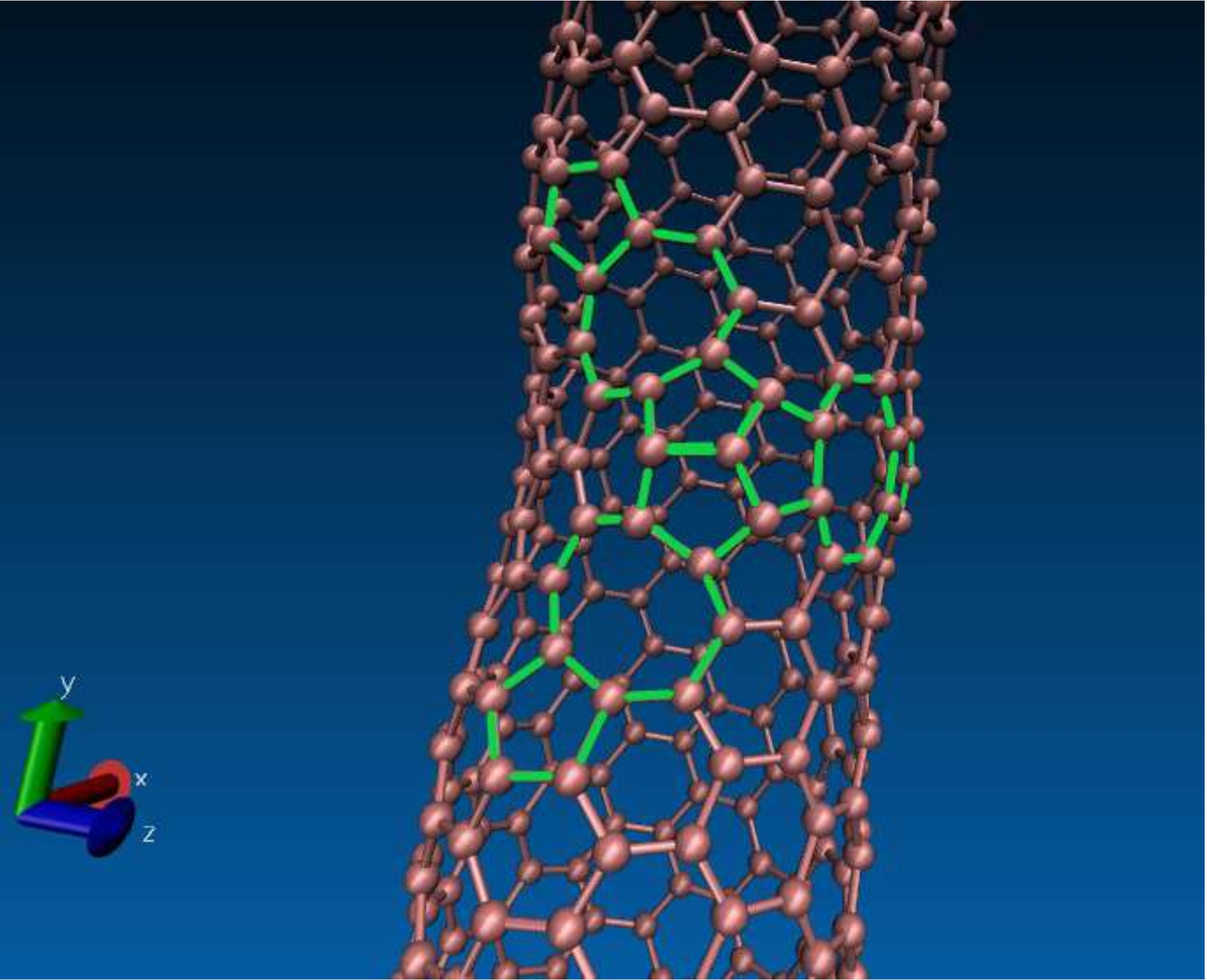}
	\caption{A $(6,8)$ nanotube with three $5-8-5$ Stone-Wales defects after relaxation using LAMMPS molecular dynamics.}
	\label{Stones}
\end{figure}

Stone-Wales defects are of significant interest due to their frequent occurrence in carbon nanotubes.
The most common Stone-Wales defect found in graphene is $5-7-7-5$, where four adjacent hexagons reconstruct themselves in two pentagons and two heptagons. 
However, the $5-8-5$ Stone-Wales defects -- corresponding to a relaxed bivacancy -- have received recent attention and are relatively less understood \cite{wang2015large,robertson2014stability,robertson2012spatial}, and therefore we use the PML method to study the defect modes in a cluster of these defects.
Three $5-8-5$ Stone-Wales defects are constructed such that they meet at a common point, as shown in Figure \ref{Stones}, motivated by the molecular dynamics simulations of similar configurations in graphene \cite{wang2015large}.

\begin{figure}[htb!]
	\centering
	\includegraphics[width=0.45\textwidth]{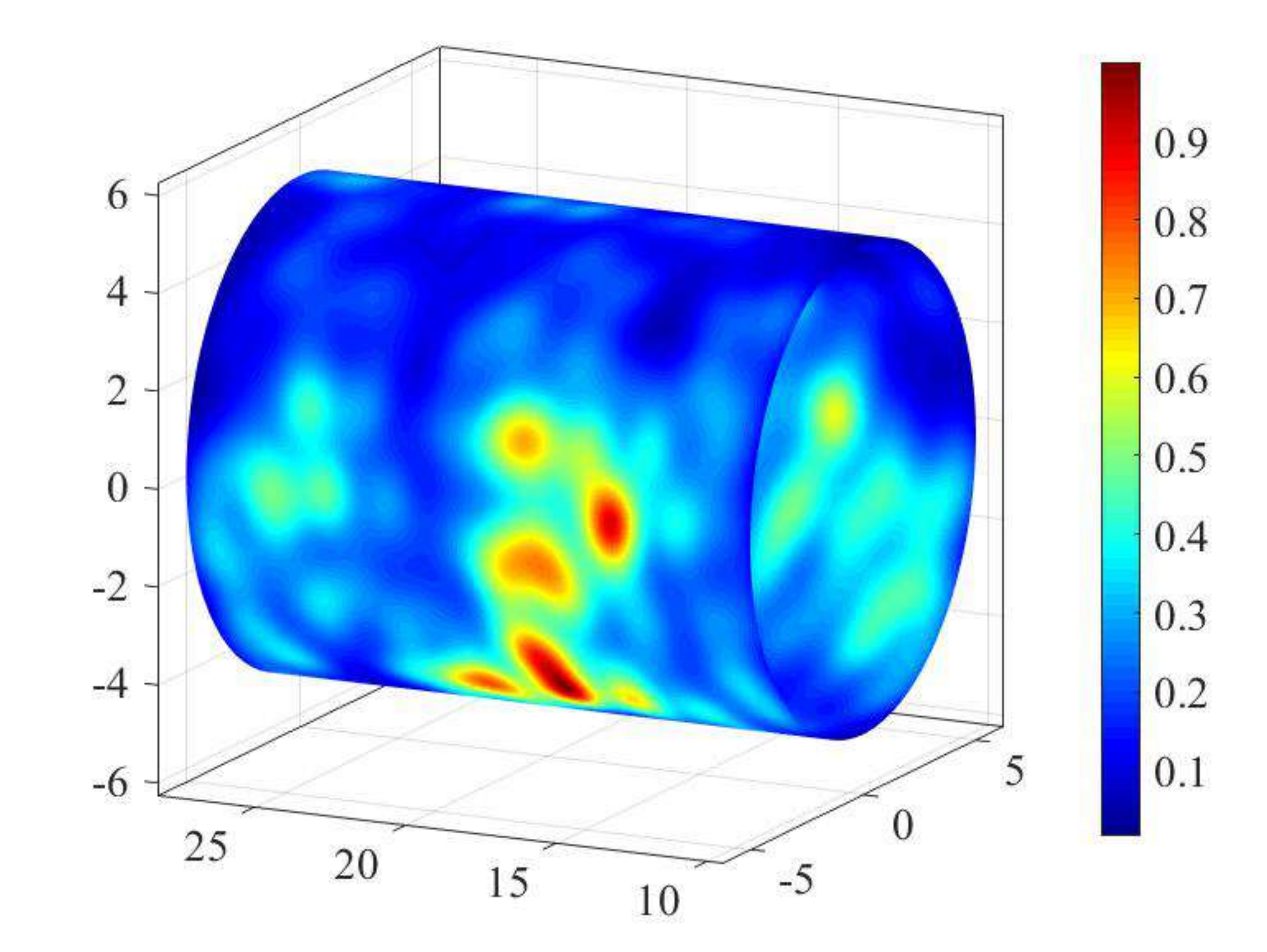}
	\includegraphics[width=0.45\textwidth]{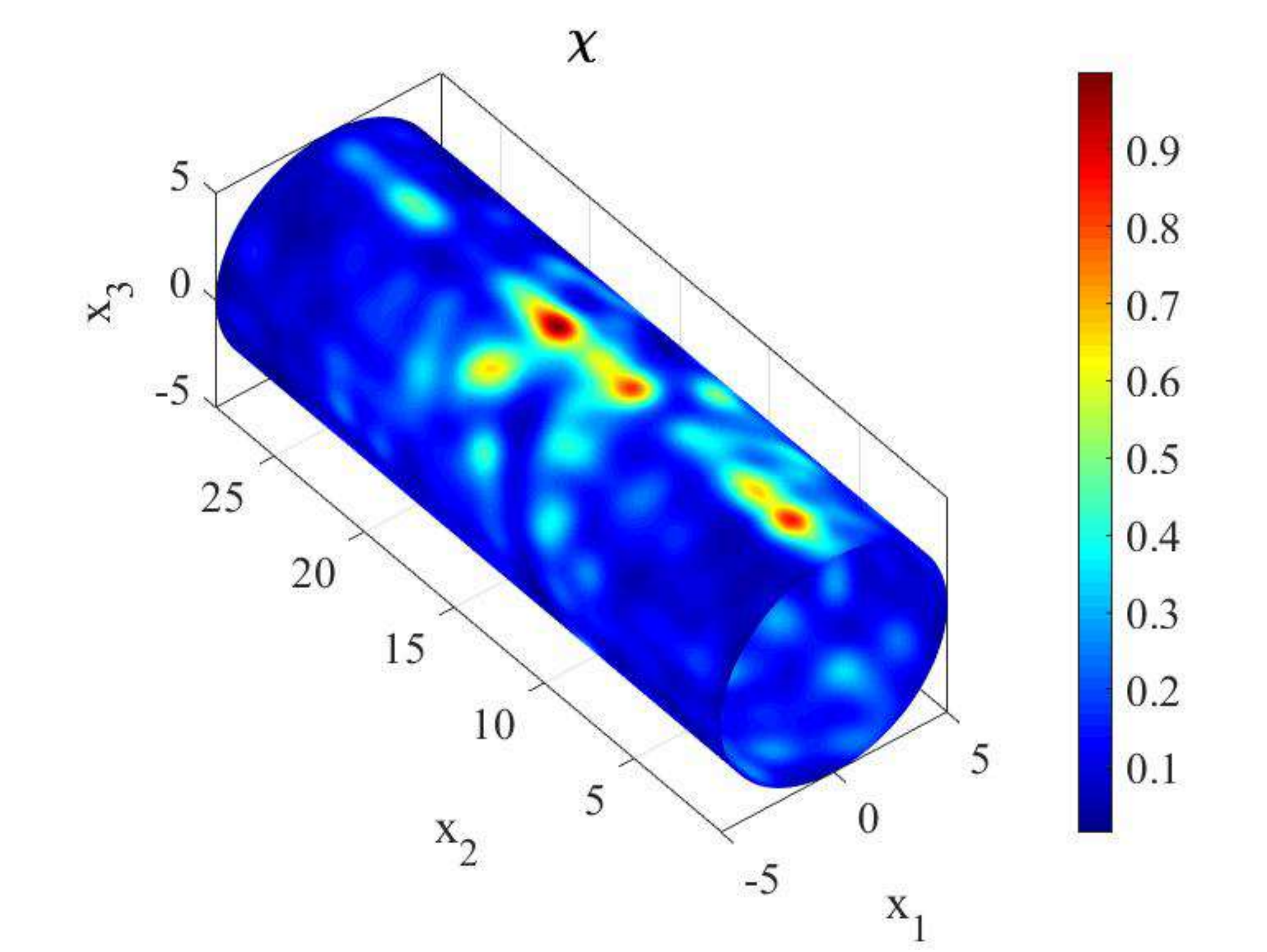}
	\caption{Defect modes for $(6,8)$ Nanotube with Stone-Wales defect. All lengths are in dimensions of $\mbox{\normalfont\AA}$.}
	\label{D-Mode}
\end{figure}

We consider a chiral $(6,8)$ nanotube.
To construct the $5-8-5$ defect cluster, we remove a pair of carbon atoms, and relax the structure using molecular dynamics simulations in LAMMPS with the microcanonical ensemble.
We then use the PML method on this relaxed configuration to find the defect modes, and some examples are shown in Figure \ref{D-Mode}.
All of these modes lie in the bandgap of the $(6,8)$ nanotube and do not correspond to any Bloch wave solution. 
These results qualitatively match with the electronic structure of a nanotube with $5-7-7-5$ Stone-wales defects in \cite{choi2000defects}.

\section{Conclusion}

In this paper, we used the symmetry-adapted framework of Objective Structures to obtain the analog in nanotubes of classical Bloch waves, and built on this to develop a perfectly-matched layer method for defects in the setting of tight binding.
The nanotube geometries that result from the deformation or the intrinsic chirality can be non-periodic, or only periodic with the use of large unit cells.
However, the symmetry-adaptation uses isometric transformations -- namely screw and rotation operations -- to describe all kinds of nanotubes, as well as transparently account for deformations.

We applied the symmetry-adapted Bloch-wave analogs to compute band structures for stretched and twisted nanotubes with different chiralities. 
The resulting band structures and density of states show the significant effect of twist on electronic properties of nanotube; torsion can observably reduce the bandgap, or even cause it to vanish in a semiconducting chiral nanotube, as well as introduce a bandgap in a conducting armchair nanotube; twisting and changes in chirality have qualitatively the same effect.
These findings agree with the analytical results provided by \cite{yang2000electronic}. 

We then studied the effects of selected defects on the electronic structure and electromechanics.
The lack of symmetry induced by the presence of the defect requires large computational domains to avoid spurious size-effects.
We therefore apply PML to obtain a computationally tractable approach by allowing us to truncate the domain without spurious boundary effects.
We apply this to study the scattering of Bloch waves at vacancies in nanotubes, and also study the interaction with twist and stretch for different chiralities.

We next examined localized bending or kinking, treated as a geometric defect, in both conducting armchair and semiconducting chiral nanotubes.
The findings agree well with the experimental observations of \cite{tom00}. 
We also compare our results with computational results that are reported in the literature \cite{chico1996quantum,LAMBIN199585,choi2000defects,zhang2009electromechanical}. 
These works have largely focused on semiconducting nanotubes, with the observation that defect modes appear in the bandgaps due to bending. 
The PML method enables us to examine the spatial structure of the defect modes in greater detail.

Finally, we examined a combination of three $5-8-5$ Stone-Wales defects in a semiconducting chiral nanotube.
Various prior works have examined the density of states (DOS) plot to find defect modes \cite{shan2005first,meunier1998energetics,chico1996quantum,LAMBIN199585}, and the PML method enabled us to probe the spatial structure of the defect modes.

In future work, the use of symmetry-adapted PML methods can be applied to the setting of the linear scaling multiple scattering (LSMS) formulation of density functional theory \cite{wang1995order}.
The LSMS approach relies heavily on the wave scattering picture of quantum mechanics, thereby providing a natural framework for the use of PML-based methods.

\section*{Research Data}
	A version of the code developed for this work is available at \url{github.com/soumyam-code/Tight-binding-nanotube}.

\begin{acknowledgments}
	We thank the National Science Foundation [CMMI MOMS 1635407], Army Research Office [W911NF-17-1-0084], Office of Naval Research [N00014-18-1-2528], and Air Force Office of Scientific Research [MURI FA9550-18-1-0095] for financial support.
	We acknowledge NSF for XSEDE computing resources provided by Pittsburgh Supercomputing Center.
\end{acknowledgments}

\bibliographystyle{alpha}
\bibliography{bib_ALL}

\newcommand{\etalchar}[1]{$^{#1}$}
\begin{thebibliography}{JHKvdZ{\etalchar{+}}05}

\bibitem[AAMW03]{avouris2003carbon}
Phaedon Avouris, Joerg Appenzeller, Richard Martel, and Shalom~J Wind.
\newblock Carbon nanotube electronics.
\newblock {\em Proceedings of the IEEE}, 91(11):1772--1784, 2003.

\bibitem[AD11]{aghaei2011symmetry}
Amin Aghaei and Kaushik Dayal.
\newblock Symmetry-adapted non-equilibrium molecular dynamics of chiral carbon
  nanotubes under tensile loading.
\newblock {\em Journal of Applied Physics}, 109(12):123501, 2011.

\bibitem[AD12]{aghaei2012tension}
Amin Aghaei and Kaushik Dayal.
\newblock Tension and twist of chiral nanotubes: torsional buckling, mechanical
  response and indicators of failure.
\newblock {\em Modelling and Simulation in Materials Science and Engineering},
  20(8):085001, 2012.

\bibitem[ADE13a]{aghaei2013anomalous}
Amin Aghaei, Kaushik Dayal, and Ryan~S Elliott.
\newblock Anomalous phonon behavior of carbon nanotubes: First-order influence
  of external load.
\newblock {\em Journal of Applied Physics}, 113(2):023503, 2013.

\bibitem[ADE13b]{aghaei2013symmetry}
Amin Aghaei, Kaushik Dayal, and Ryan~S Elliott.
\newblock Symmetry-adapted phonon analysis of nanotubes.
\newblock {\em Journal of the Mechanics and Physics of Solids}, 61(2):557--578,
  2013.

\bibitem[AO10]{ariza2010discrete}
MP~Ariza and M~Ortiz.
\newblock Discrete dislocations in graphene.
\newblock {\em Journal of the Mechanics and Physics of Solids}, 58(5):710--734,
  2010.

\bibitem[ASMO12]{ariza2012stacking}
MP~Ariza, R~Serrano, JP~Mendez, and M~Ortiz.
\newblock Stacking faults and partial dislocations in graphene.
\newblock {\em Philosophical Magazine}, 92(16):2004--2021, 2012.

\bibitem[AWHS17]{ahmadpoor2017thermal}
Fatemeh Ahmadpoor, Peng Wang, Rui Huang, and Pradeep Sharma.
\newblock Thermal fluctuations and effective bending stiffness of elastic thin
  sheets and graphene: A nonlinear analysis.
\newblock {\em Journal of the Mechanics and Physics of Solids}, 107:294--319,
  2017.

\bibitem[BC04]{basu2004perfectly}
Ushnish Basu and Anil~K Chopra.
\newblock Perfectly matched layers for transient elastodynamics of unbounded
  domains.
\newblock {\em International Journal for Numerical Methods in Engineering},
  59(8):1039--1074, 2004.

\bibitem[BDJR05]{bandaru2005novel}
Prabhakar~R Bandaru, Chiara Daraio, Sungho Jin, and Apparao~M Rao.
\newblock Novel electrical switching behaviour and logic in carbon nanotube
  y-junctions.
\newblock {\em Nature materials}, 4(9):663--666, 2005.

\bibitem[Ber94]{BERENGER1994185}
Jean-Pierre Berenger.
\newblock A perfectly matched layer for the absorption of electromagnetic
  waves.
\newblock {\em Journal of Computational Physics}, 114(2):185 -- 200, 1994.

\bibitem[BGG03]{bradley2003flexible}
Keith Bradley, Jean-Christophe~P Gabriel, and George Gr{\"u}ner.
\newblock Flexible nanotube electronics.
\newblock {\em Nano Letters}, 3(10):1353--1355, 2003.

\bibitem[BIC{\etalchar{+}}05]{bekyarova2005electronic}
Elena Bekyarova, Mikhail~E Itkis, Nelson Cabrera, Bin Zhao, Aiping Yu, Junbo
  Gao, and Robert~C Haddon.
\newblock Electronic properties of single-walled carbon nanotube networks.
\newblock {\em Journal of the American Chemical Society}, 127(16):5990--5995,
  2005.

\bibitem[CBLC96]{chico1996quantum}
Leonor Chico, Lorin~X Benedict, Steven~G Louie, and Marvin~L Cohen.
\newblock Quantum conductance of carbon nanotubes with defects.
\newblock {\em Physical Review B}, 54(4):2600, 1996.

\bibitem[CBR07]{charlier2007electronic}
Jean-Christophe Charlier, Xavier Blase, and Stephan Roche.
\newblock Electronic and transport properties of nanotubes.
\newblock {\em Reviews of modern physics}, 79(2):677, 2007.

\bibitem[CCC{\etalchar{+}}19]{cao2019review}
Yu~Cao, Sen Cong, Xuan Cao, Fanqi Wu, Qingzhou Liu, Moh~R Amer, and Chongwu
  Zhou.
\newblock Review of electronics based on single-walled carbon nanotubes.
\newblock In {\em Single-Walled Carbon Nanotubes}, pages 189--224. Springer,
  2019.

\bibitem[CILC00]{choi2000defects}
Hyoung~Joon Choi, Jisoon Ihm, Steven~G Louie, and Marvin~L Cohen.
\newblock Defects, quasibound states, and quantum conductance in metallic
  carbon nanotubes.
\newblock {\em Physical Review Letters}, 84(13):2917, 2000.

\bibitem[CLRA07]{Chen2007228}
Zhihong Chen, Yu-Ming Lin, Michael~J. Rooks, and Phaedon Avouris.
\newblock Graphene nano-ribbon electronics.
\newblock {\em Physica E: Low-dimensional Systems and Nanostructures},
  40(2):228 -- 232, 2007.
\newblock International Symposium on Nanometer-Scale Quantum Physics.

\bibitem[CW94]{chew19943d}
Weng~Cho Chew and William~H Weedon.
\newblock A 3d perfectly matched medium from modified maxwell's equations with
  stretched coordinates.
\newblock {\em Microwave and optical technology letters}, 7(13):599--604, 1994.

\bibitem[DEJ]{dayalFormulae}
Kaushik Dayal, R.~S. Elliott, and Richard~D James.
\newblock Formulas for objective structures.
\newblock {\em preprint}.

\bibitem[Dek18]{dekker2018we}
Cees Dekker.
\newblock How we made the carbon nanotube transistor.
\newblock {\em Nature Electronics}, 1(9):518--518, 2018.

\bibitem[DJ07]{dumitricua2007objective}
Traian Dumitric{\u{a}} and Richard~D James.
\newblock Objective molecular dynamics.
\newblock {\em Journal of the Mechanics and Physics of Solids},
  55(10):2206--2236, 2007.

\bibitem[DJ10]{dayal2010nonequilibrium}
Kaushik Dayal and Richard~D James.
\newblock Nonequilibrium molecular dynamics for bulk materials and
  nanostructures.
\newblock {\em Journal of the Mechanics and Physics of Solids}, 58(2):145--163,
  2010.

\bibitem[GS17]{ghosh2017sparc}
Swarnava Ghosh and Phanish Suryanarayana.
\newblock Sparc: Accurate and efficient finite-difference formulation and
  parallel implementation of density functional theory: Isolated clusters.
\newblock {\em Computer Physics Communications}, 212:189--204, 2017.

\bibitem[HPK{\etalchar{+}}11]{huang2011electronic}
Mingyuan Huang, Tod~A Pascal, Hyungjun Kim, William~A Goddard~III, and Julia~R
  Greer.
\newblock Electronic- mechanical coupling in graphene from in situ
  nanoindentation experiments and multiscale atomistic simulations.
\newblock {\em Nano letters}, 11(3):1241--1246, 2011.

\bibitem[HSP69]{hehre1969self}
Warren~J Hehre, Robert~F Stewart, and John~A Pople.
\newblock self-consistent molecular-orbital methods. i. use of gaussian
  expansions of slater-type atomic orbitals.
\newblock {\em The Journal of Chemical Physics}, 51(6):2657--2664, 1969.

\bibitem[Jam06]{james2006objective}
Richard~D James.
\newblock Objective structures.
\newblock {\em Journal of the Mechanics and Physics of Solids},
  54(11):2354--2390, 2006.

\bibitem[JHKvdZ{\etalchar{+}}05]{PhysRevLett.94.156802}
P.~Jarillo-Herrero, J.~Kong, H.~S.~J. van~der Zant, C.~Dekker, L.~P.
  Kouwenhoven, and S.~De~Franceschi.
\newblock Electronic transport spectroscopy of carbon nanotubes in a magnetic
  field.
\newblock {\em Phys. Rev. Lett.}, 94:156802, Apr 2005.

\bibitem[JK09]{javey2009carbon}
Ali Javey and Jing Kong.
\newblock {\em Carbon nanotube electronics}.
\newblock Springer Science \& Business Media, 2009.

\bibitem[KPK{\etalchar{+}}03]{kramberger2003assignment}
Ch~Kramberger, R~Pfeiffer, H~Kuzmany, V~Z{\'o}lyomi, and J~K{\"u}rti.
\newblock Assignment of chiral vectors in carbon nanotubes.
\newblock {\em Physical Review B}, 68(23):235404, 2003.

\bibitem[KS15]{kerszberg2015ab}
Nicolas Kerszberg and Phanish Suryanarayana.
\newblock Ab initio strain engineering of graphene: opening bandgaps up to 1
  ev.
\newblock {\em RSC Advances}, 5(54):43810--43814, 2015.

\bibitem[KS20]{kumar2020bending}
Shashikant Kumar and Phanish Suryanarayana.
\newblock Bending moduli for thirty-two select atomic monolayers from first
  principles.
\newblock {\em arXiv preprint arXiv:2003.10936}, 2020.

\bibitem[LDS{\etalchar{+}}05]{li2005continuum}
Zhiling Li, Prasad Dharap, Pradeep Sharma, Satish Nagarajaiah, and Boris~I
  Yakobson.
\newblock Continuum field model of defect formation in carbon nanotubes.
\newblock {\em Journal of applied physics}, 97(7):074303, 2005.

\bibitem[LFV{\etalchar{+}}95]{LAMBIN199585}
Ph. Lambin, A.~Fonseca, J.P. Vigneron, J.B. Nagy, and A.A. Lucas.
\newblock Structural and electronic properties of bent carbon nanotubes.
\newblock {\em Chemical Physics Letters}, 245(1):85 -- 89, 1995.

\bibitem[Mar04]{martin2004electronic}
Richard~M Martin.
\newblock {\em Electronic structure: basic theory and practical methods}.
\newblock Cambridge university press, 2004.

\bibitem[MFP02]{mceuen2002single}
Paul~L McEuen, Michael~S Fuhrer, and Hongkun Park.
\newblock Single-walled carbon nanotube electronics.
\newblock {\em IEEE transactions on nanotechnology}, 1(1):78--85, 2002.

\bibitem[MHL98]{meunier1998energetics}
Vincent Meunier, Luc Henrard, and Ph~Lambin.
\newblock Energetics of bent carbon nanotubes.
\newblock {\em Physical Review B}, 57(4):2586, 1998.

\bibitem[NDJD14]{nikiforov2014tight}
I~Nikiforov, E~Dontsova, RD~James, and T~Dumitric{\u{a}}.
\newblock Tight-binding theory of graphene bending.
\newblock {\em Physical Review B}, 89(15):155437, 2014.

\bibitem[PD16]{Pourmatin2016115}
Hossein Pourmatin and Kaushik Dayal.
\newblock Multiscale real-space quantum-mechanical tight-binding calculations
  of electronic structure in crystals with defects using perfectly matched
  layers.
\newblock {\em Journal of Computational Physics}, 323:115 -- 125, 2016.

\bibitem[Pop04]{Popov_2004}
Valentin~N Popov.
\newblock Curvature effects on the structural, electronic and optical
  properties of isolated single-walled carbon nanotubes within a
  symmetry-adapted non-orthogonal tight-binding model.
\newblock {\em New Journal of Physics}, 6:17--17, feb 2004.

\bibitem[PP99]{pickard1999extrapolative}
CJ~Pickard and MC~Payne.
\newblock Extrapolative approaches to brillouin-zone integration.
\newblock {\em Physical Review B}, 59(7):4685, 1999.

\bibitem[PRL{\etalchar{+}}00]{papadopoulos2000electronic}
C~Papadopoulos, A~Rakitin, J~Li, AS~Vedeneev, and JM~Xu.
\newblock Electronic transport in y-junction carbon nanotubes.
\newblock {\em Physical review letters}, 85(16):3476, 2000.

\bibitem[RAW{\etalchar{+}}12]{robertson2012spatial}
Alex~W Robertson, Christopher~S Allen, Yimin~A Wu, Kuang He, Jaco Olivier, Jan
  Neethling, Angus~I Kirkland, and Jamie~H Warner.
\newblock Spatial control of defect creation in graphene at the nanoscale.
\newblock {\em Nature communications}, 3:1144, 2012.

\bibitem[RLH{\etalchar{+}}14]{robertson2014stability}
Alex~W Robertson, Gun-Do Lee, Kuang He, Euijoon Yoon, Angus~I Kirkland, and
  Jamie~H Warner.
\newblock Stability and dynamics of the tetravacancy in graphene.
\newblock {\em Nano letters}, 14(3):1634--1642, 2014.

\bibitem[SLPC05]{shan2005first}
Bin Shan, Gregory~W Lakatos, Shu Peng, and Kyeongjae Cho.
\newblock First-principles study of band-gap change in deformed nanotubes.
\newblock {\em Applied Physics Letters}, 87(17):173109, 2005.

\bibitem[SRT{\etalchar{+}}19]{sahalianov2019straintronics}
Ihor~Yu Sahalianov, Taras~M Radchenko, Valentyn~A Tatarenko, Gianaurelio
  Cuniberti, and Yuriy~I Prylutskyy.
\newblock Straintronics in graphene: Extra large electronic band gap induced by
  tensile and shear strains.
\newblock {\em Journal of Applied Physics}, 126(5):054302, 2019.

\bibitem[TM11]{tadmor2011modeling}
Ellad~B Tadmor and Ronald~E Miller.
\newblock {\em Modeling materials: continuum, atomistic and multiscale
  techniques}.
\newblock Cambridge University Press, 2011.

\bibitem[Tom00]{tom00}
Thomas~W. Tombler.
\newblock Reversible electromechanical characteristics of carbon nanotubes
  under local-probe manipulation.
\newblock {\em Nature}, 5, 2000.

\bibitem[TWT19]{tan2019dislocation}
Anne Marie~Z Tan, Christopher Woodward, and Dallas~R Trinkle.
\newblock Dislocation core structures in ni-based superalloys computed using a
  density functional theory based flexible boundary condition approach.
\newblock {\em Physical Review Materials}, 3(3):033609, 2019.

\bibitem[VAKCC03]{PhysRevB.67.161401}
M.~Verissimo-Alves, Belita Koiller, H.~Chacham, and R.~B. Capaz.
\newblock Electromechanical effects in carbon nanotubes: \textit{Ab initio} and
  analytical tight-binding calculations.
\newblock {\em Phys. Rev. B}, 67:161401, Apr 2003.

\bibitem[Wal47]{wallace1947band}
Philip~Richard Wallace.
\newblock The band theory of graphite.
\newblock {\em Physical review}, 71(9):622, 1947.

\bibitem[WSS{\etalchar{+}}95]{wang1995order}
Yang Wang, GM~Stocks, WA~Shelton, DMC Nicholson, Z~Szotek, and WM~Temmerman.
\newblock Order-n multiple scattering approach to electronic structure
  calculations.
\newblock {\em Physical review letters}, 75(15):2867, 1995.

\bibitem[WTTJ13]{wang2013carbon}
Chuan Wang, Kuniharu Takei, Toshitake Takahashi, and Ali Javey.
\newblock Carbon nanotube electronics--moving forward.
\newblock {\em Chemical Society Reviews}, 42(7):2592--2609, 2013.

\bibitem[WYY{\etalchar{+}}15]{wang2015large}
Shuaiwei Wang, Baocheng Yang, Jinyun Yuan, Yubing Si, and Houyang Chen.
\newblock Large-scale molecular simulations on the mechanical response and
  failure behavior of a defective graphene: Cases of 5--8--5 defects.
\newblock {\em Scientific reports}, 5, 2015.

\bibitem[YH00]{yang2000electronic}
Liu Yang and Jie Han.
\newblock Electronic structure of deformed carbon nanotubes.
\newblock {\em Physical Review Letters}, 85(1):154, 2000.

\bibitem[ZAGS17]{zelisko2017determining}
Matthew Zelisko, Fatemeh Ahmadpoor, Huajian Gao, and Pradeep Sharma.
\newblock Determining the gaussian modulus and edge properties of 2d materials:
  From graphene to lipid bilayers.
\newblock {\em Physical review letters}, 119(6):068002, 2017.

\bibitem[ZD08]{zhang2008elasticity}
D-B Zhang and T~Dumitric{\u{a}}.
\newblock Elasticity of ideal single-walled carbon nanotubes via
  symmetry-adapted tight-binding objective modeling.
\newblock {\em Applied physics letters}, 93(3):031919, 2008.

\bibitem[ZHD08]{zhang2008stability}
D-B Zhang, M~Hua, and T~Dumitric{\u{a}}.
\newblock Stability of polycrystalline and wurtzite si nanowires via
  symmetry-adapted tight-binding objective molecular dynamics.
\newblock {\em The Journal of chemical physics}, 128(8):084104, 2008.

\bibitem[ZJD09a]{zhang2009dislocation}
D-B Zhang, Richard~D James, and T~Dumitric{\u{a}}.
\newblock Dislocation onset and nearly axial glide in carbon nanotubes under
  torsion.
\newblock {\em The Journal of chemical physics}, 130:071101, 2009.

\bibitem[ZJD09b]{zhang2009electromechanical}
D-B Zhang, Richard~D James, and T~Dumitric{\u{a}}.
\newblock Electromechanical characterization of carbon nanotubes in torsion via
  symmetry adapted tight-binding objective molecular dynamics.
\newblock {\em Physical Review B}, 80(11):115418, 2009.

\bibitem[ZJSY06]{zhang2006atomistic}
X~Zhang, K~Jiao, P~Sharma, and BI~Yakobson.
\newblock An atomistic and non-classical continuum field theoretic perspective
  of elastic interactions between defects (force dipoles) of various symmetries
  and application to graphene.
\newblock {\em Journal of the Mechanics and Physics of Solids},
  54(11):2304--2329, 2006.

\bibitem[ZPHL04]{zhao2004electronic}
Jijun Zhao, Hyoungki Park, Jie Han, and Jian~Ping Lu.
\newblock Electronic properties of carbon nanotubes with covalent sidewall
  functionalization.
\newblock {\em The Journal of Physical Chemistry B}, 108(14):4227--4230, 2004.

\end{thebibliography}

\end{document}